\PassOptionsToPackage{expansion=false}{microtype}
\PassOptionsToPackage{table}{xcolor}

\documentclass[letterpaper,twocolumn,10pt]{article}
\usepackage{usenix2019_v3}

\usepackage{amsmath}
\usepackage{amssymb}
\usepackage{amsthm}
\usepackage{makecell}
\usepackage{booktabs}
\usepackage{pifont}
\usepackage{subcaption}
\usepackage{xurl}
\usepackage{xspace}
\usepackage{listings}
\usepackage{graphicx}
\usepackage{textcomp}
\usepackage{tabularx}
\usepackage{float}
\graphicspath{{./}}
\usepackage{tikz}
\usetikzlibrary{arrows.meta,positioning,fit,backgrounds,calc,shapes.geometric}
\usepackage{enumitem}
\setlist{nosep,leftmargin=*}
\providecommand{\Description}[1]{}  % figure alt-text, acmart only
\providecommand{\keywords}[1]{}     % keyword list, acmart only

\theoremstyle{definition}
\newtheorem{definition}{Definition}

\newcommand{\etal}{et~al.\xspace}

\date{}
\title{\Large \bf Self-State Attacks on Self-Hosted AI Agents:\\
  How Far Can OS Defenses Go?}

\author{
{\rm Yimeng Chen$^{1}$, Nathana\"el Denis$^{1}$, Roberto Di Pietro$^{1}$, J\"urgen Schmidhuber$^{1,2}$}\\
$^{1}$Center of Excellence for Generative AI, King Abdullah University of Science and Technology\\
$^{2}$The Swiss AI Lab, IDSIA-USI/SUPSI\\
\{yimeng.chen, nathanael.denis, roberto.dipietro, juergen.schmidhuber\}@kaust.edu.sa
}

\begin{document}

\maketitle

\begin{abstract}
    Self-hosted AI agents maintain persistent memory, instructions, and configuration that influence their future behavior. If an agent is compromised, an attacker can exploit the agent's legitimate write permissions to corrupt this self-state, making malicious and benign updates difficult to distinguish at the operating system (OS) level. We investigate how far existing OS mechanisms can prevent, detect, and recover from such self-state attacks. We formalize an attack space and evaluate representative OS defenses using four agent workloads and a Linux telemetry pipeline. Our results show a consistent limitation across defense dimensions. File-level controls either leave alternative mutation paths open or, when complete over the tested operations, also block corresponding legitimate updates. Detectors flag a substantial part of legitimate activity, while more selective methods cover only part of the attack space. Finally, protected backups successfully restore corrupted state, but require a trusted recovery point and may incur rollback cost. Overall, our results show that the main limitation is not OS observability. Indeed, the OS can enforce, observe, attribute, and recover self-state changes. Yet, generic OS defenses lack the decision context needed to combine broad operation coverage with selective decisions. Effective protection therefore requires self-state-aware mechanisms that exploit additional context beyond generic file and syscall behavior.
\end{abstract}

% USENIX papers carry no keyword list; retained here for reference only.

%% ===========================================================
%% SECTION 1: Introduction
%% ===========================================================
\section{Introduction}
\label{sec:intro}

LLM-based agents increasingly execute tasks, invoke tools, and retain state
across interactions. Self-hosted systems place these capabilities on the
user's machine so that an agent can work inside local codebases, documents,
and workflows through ordinary shell and filesystem interfaces.
Here, self-hosted refers to deployment of the agent runtime and
persistent state on a user-controlled host. Thus, the underlying model may be local
or remote. This design is used by systems such as Claude Code, Open Interpreter,
Aider, and OpenClaw~\cite{claudecode,openinterpreter,aider,openclaw}, alongside other
self-hosted agent frameworks~\cite{opencode,nanoclaw,hermesagent}.

Self-hosted agents are characterized by persistent state that shapes later
behavior: memory accumulated across sessions, instructions that define persona
and policy, and configuration that controls runtime behavior. These state roles
need not share a filename or mutability policy. The agent must update some of
them, such as episodic memory, during normal operation. Conversely, other roles may be
updated only under particular workloads or operator-approved workflows. Thus,
a deployment retains an authorized write surface over at least part of the
state that governs future decisions.

This selective self-mutability creates a distinctive class of threats. Once
the agent's decision process is semantically compromised, for example, through
an indirect prompt injection~\cite{greshake2023}, an authorized self-state
write becomes a channel for persistent corruption. We call this class
\emph{self-state attacks} (\S\ref{sec:selfstate}). This setting resembles the
classical confused deputy problem~\cite{hardy1988,confusedpilot2024}, where a
privileged component is tricked into misusing its authority. In LLM-based
agents, however, this issue is amplified by the flexibility of natural
language and the agent's ability to persist changes over time. Real-world
incidents and recent studies further highlight the severity of these
risks~\cite{promptware2026,openclawsecurity2026,habler2026memorypoisoning}.

Many agent security defenses operate at the semantic or agent runtime layer,
using LLM-as-judge methods~\cite{zheng2023llmjudge,rjudge2024}, guard
agents~\cite{guardagent2024}, or explicit policies to reject malicious actions
before execution. These defenses can add intent and content context that is not
present in the operation-level metadata evaluated here, but their decisions
occur before the resulting action reaches the host. Adaptive attacks show that
this semantic boundary remains an active area of
study~\cite{adaptiveattacks2025}. This leaves the host as a complementary
defense surface once an action is issued.

\begin{figure*}[t]
    \centering
    \resizebox{\textwidth}{!}{%% Figure 1: Cross-boundary vs self-state attacks (TikZ).
%% Two-panel side-by-side layout (use inside figure*, not figure).
%% Requires tikz libs: arrows.meta, positioning, fit, backgrounds, calc.

\begin{tikzpicture}[
  font=\footnotesize,
  >={Stealth[length=4.5pt,width=3.5pt]},
  every node/.style={inner sep=2pt},
  agent/.style={
    rounded corners=2pt, draw=black!70, fill=black!5,
    minimum width=18mm, minimum height=8mm,
    align=center, font=\footnotesize\bfseries
  },
  resource/.style={
    rounded corners=1pt, draw=black!55, fill=white,
    minimum width=16mm, minimum height=5.2mm,
    align=center, font=\footnotesize
  },
  selfres/.style={
    rounded corners=1pt, draw=black!55, fill=black!3,
    minimum width=24mm, minimum height=5.2mm,
    align=center, font=\footnotesize
  },
  inject/.style={
    draw=black!55, fill=white, rounded corners=1pt,
    minimum width=15mm, minimum height=8mm,
    align=center, font=\footnotesize
  },
  boundary/.style={
    draw=black!55, dashed, rounded corners=3pt, thick,
    inner sep=2.5mm
  },
  dangerbox/.style={
    draw=red!65!black, dashed, rounded corners=3pt, thick,
    inner sep=2.5mm
  },
  opflow/.style={->, thick, black!75},
  attack/.style={->, thick, red!65!black},
  legit/.style={->, thick, black!60, densely dashed},
  paneltitle/.style={font=\footnotesize\bfseries, anchor=west},
  panelformula/.style={font=\footnotesize, anchor=west},
]

%% =========================================================
%% Panel (a): Cross-boundary attack -- LEFT
%% =========================================================
\begin{scope}[local bounding box=panelA]

  \node[inject] (injA) at (0,0) {Prompt\\injection};
  \node[agent, right=10mm of injA] (agentA) {Compromised\\agent};

  \node[resource, right=20mm of agentA, yshift=10mm]  (filesA) {User files};
  \node[resource, right=20mm of agentA]               (netA)   {Network};
  \node[resource, right=20mm of agentA, yshift=-10mm] (hostA)  {Host proc.};

  \draw[opflow] (injA) -- (agentA);

  % Cross-boundary operations are visibly malicious.
  \draw[attack] (agentA) -- node[above, sloped, font=\scriptsize]{\texttt{rm -rf}} (filesA);
  \draw[attack] (agentA) -- node[above, font=\scriptsize]{exfiltrate} (netA);
  \draw[attack] (agentA) -- node[below, sloped, font=\scriptsize]{code exec} (hostA);

  \begin{pgfonlayer}{background}
    \node[boundary, fit=(filesA)(netA)(hostA)] (isoA) {};
  \end{pgfonlayer}

  \node[font=\scriptsize, anchor=north east, text=black!60] at (isoA.south east)
    {OS isolation (sandbox, seccomp, netns)};

  \node[paneltitle] at ($(injA.north west)+(0,9mm)$) (hdrA)
    {(a) Cross-boundary attack:};
  \node[panelformula, right=2mm of hdrA]
    {$\pi_{\mathcal O}(a)\notin U_{\mathcal O}$};

\end{scope}

%% =========================================================
%% Panel (b): Self-state attack -- RIGHT
%% =========================================================
\begin{scope}[local bounding box=panelB, shift={($(panelA.east)+(20mm,0)$)}]

  \node[inject] (injB) at (0,0) {Prompt\\injection};
  \node[agent, right=10mm of injB] (agentB) {Compromised\\agent};

  % Extra space leaves room for the operation-equivalence annotation.
  \node[selfres, right=32mm of agentB, yshift=10mm]  (memB)  {\texttt{MEMORY.md}};
  \node[selfres, right=32mm of agentB]               (idB)   {\texttt{SOUL.md}/\texttt{AGENTS.md}};
  \node[selfres, right=32mm of agentB, yshift=-10mm] (cfgB)  {\texttt{openclaw.json}/\texttt{.env}};

  \draw[opflow] (injB) -- (agentB);

  % IMPORTANT:
  % At the OS-visible boundary these self-state writes are not visually
  % classified as malicious. Draw only the legitimate/authorized operation
  % style; the text and formula explain that an attacker-induced write can
  % have the same projection.
  \draw[legit] (agentB.east) to[out=28,in=180] (memB.west);
  \draw[legit] (agentB.east) -- (idB.west);
  \draw[legit] (agentB.east) to[out=-28,in=180] (cfgB.west);

  % Operation-equivalence annotation in open space, with white backing.
  \node[
    font=\scriptsize\itshape,
    anchor=south,
    inner xsep=1.5pt,
    inner ysep=0.8pt,
    fill=white,
    text=black!75
  ] at ($(agentB.east)!0.56!(idB.west)+(0,2.2mm)$)
    {same OS-visible op};

  \begin{pgfonlayer}{background}
    \node[dangerbox, fit=(memB)(idB)(cfgB)] (isoB) {};
  \end{pgfonlayer}

  \node[font=\scriptsize, anchor=north east, text=red!65!black] at (isoB.south east)
    {Metadata-only policy cannot separate the overlapping pair};

  \node[paneltitle] at ($(injB.north west)+(0,9mm)$) (hdrB)
    {(b) Self-state attack:};
  \node[panelformula, right=2mm of hdrB, text=red!65!black]
    {$\pi_{\mathcal O}(a)=\pi_{\mathcal O}(\ell)\in U_{\mathcal O}$};

\end{scope}

%% =========================================================
%% Shared legend for BOTH panels
%% =========================================================
\coordinate (legendMid) at ($(panelA.south east)!0.5!(panelB.south west)+(0,-5mm)$);

\draw[legit] ($(legendMid)+(-22mm,0)$) -- ++(8mm,0)
  node[anchor=west, font=\scriptsize, text=black!70]{legit / authorized operation};

\draw[attack] ($(legendMid)+(18mm,0)$) -- ++(8mm,0)
  node[anchor=west, font=\scriptsize, text=red!65!black]{malicious operation};

%% Light vertical divider between the two panels
\draw[black!25, very thin]
  ($(panelA.north east)!0.5!(panelB.north west)+(0,3mm)$)
  -- ($(panelA.south east)!0.5!(panelB.south west)+(0,-3mm)$);

\end{tikzpicture}}
    \Description{Two-panel diagram contrasting cross-boundary and self-state attacks. Panel (a) shows a compromised agent requesting operations outside a configured isolation policy's authorized scope. Panel (b) shows an attacker-induced and a legitimate update to the agent's own memory, instruction, or configuration converging to the same operation-level policy input. A metadata-only policy cannot separate that pair.}
    \caption{Cross-boundary vs.\ self-state attacks. For a fixed operation-level observation model $\mathcal{O}$, $U_{\mathcal{O}}$ is the policy's authorized input set, $a$ is an attacker-induced operation, and $\ell$ is a legitimate update. (a) Isolation rejects an operation whose projected input falls outside $U_{\mathcal{O}}$. (b) A projection-overlapping self-state attack and a legitimate update share the same authorized input, so a policy restricted to that input assigns them the same decision.}
    \label{fig:self-state-attack}
\end{figure*}
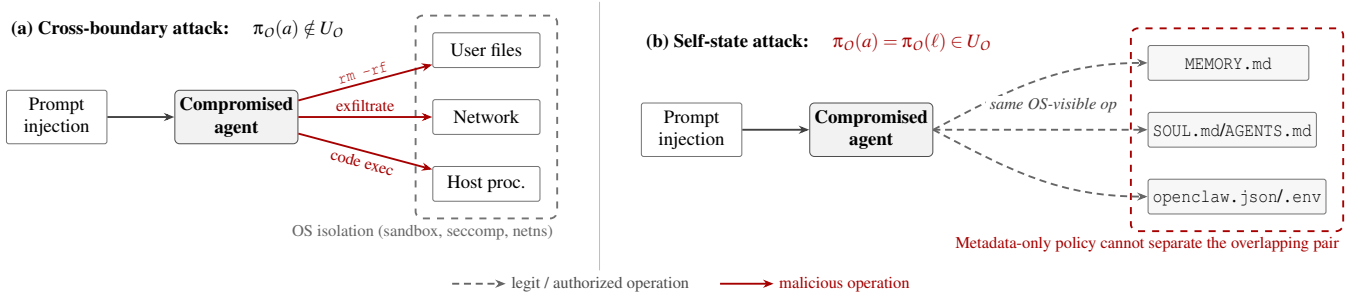

The operating system (OS) is the last deterministic enforcement point before
any self-state corruption commits to disk, since every such corruption must
eventually issue a concrete syscall, such as a write, rename, unlink, or mode
change. Yet, mediation alone does not create a selective boundary: legitimate
and updates induced by the attacker may use the same principal, target, and operation (Figure~\ref{fig:self-state-attack}). It therefore remains unclear how far OS mechanisms can defend an agent's own
writable state.

To resolve this uncertainty, we organize self-state attacks along four axes and identify the decision requirements they place on OS prevention, detection, and recovery. We then construct a benchmark whose attack catalog systematically covers mutations of memory, instructions, and configuration. Its primary catalog contains 23 structural cells. Separately, four contrastive workloads covering coding, knowledge work, operations, and general assistance exercise different patterns of legitimate self-state activity. A Linux measurement pipeline combines inotify, fanotify, auditd, eBPF, and Sysdig system-call capture to reconstruct the subject, process lineage, operation, object, and outcome of each state change, and exports the resulting evidence in the native forms consumed by representative prevention, detection, provenance, and recovery mechanisms.
% The normalized provenance graph separately supports attribution and serves as input to the graph-based detector.

The evaluation reveals a common limitation across prevention, detection, and recovery: generic OS mechanisms provide useful control and evidence, but do not by themselves support selective treatment of malicious and legitimate self-state activity.
Prevention mechanisms may either leave alternative mutation paths open or reject corresponding authorized operations. Detection exhibits a coverage-specificity tradeoff: broad methods falsely reject a substantial part of natural activity, while more selective methods cover only part of the attack space. This pattern persists across baselines, and workload conditioning provides only modest improvement. In addition, protected backups survive same-principal destruction and restore all tested corruptions, but selecting a trustworthy recovery point remains unresolved, and rollback exposure varies by workload. Overall, the results reveal a defense-design limitation rather than an observability issue. This motivates self-state-specific OS defenses that introduce state-aware decision context and protected recovery points, as discussed in Section~\ref{sec:discussion}.

\textbf{Contributions.} Our contributions can be summarized as:
\begin{itemize}
    \item We define self-state attacks as corruption of an agent's own persistent, behavior-governing state, organize them along four axes, and map OS prevention, detection, and recovery capabilities to the specific decisions each dimension must make.
    \item We construct a benchmark that combines systematic attack selection across memory, instructions, and configuration with four contrastive workloads, supported by five-source Linux telemetry collection using inotify, fanotify, auditd, eBPF, and Sysdig capture.
    \item We evaluate representative OS mechanisms across prevention, detection, and recovery, demonstrating concrete but scoped value in enforcement, attribution, and protected recovery while exposing the limitations of generic file and syscall baselines on self-state activity.
\end{itemize}

The remainder of the paper is organized as follows. Section~\ref{sec:selfstate}
formalizes self-state attacks and introduces the four-axis attack space.
Section~\ref{sec:gaps} derives the decision requirements for OS prevention,
detection, and recovery. Section~\ref{sec:experiments} describes the experimental
methodology, and Section~\ref{sec:results} reports the empirical results.
Section~\ref{sec:related} positions our work against the state of the art, and
Section~\ref{sec:conclusion} concludes.

%% ===========================================================
%% SECTION 2: Self-State Attacks
%% ===========================================================

\section{Self-State Attacks}
\label{sec:selfstate}

We first define self-state by its functional roles and separate those roles
from their platform-specific file bindings. We then state the attacker and
trust assumptions, define self-state attacks, and systematically categorize the different forms that self-state attacks can take.

\begin{table}[t]
\caption{Logical roles of agent self-state.}
\label{tab:self-state-roles}
\centering
\footnotesize
\setlength{\tabcolsep}{3pt}
\renewcommand{\arraystretch}{1.08}
\begin{tabularx}{\columnwidth}{@{}p{.17\columnwidth}p{.27\columnwidth}X@{}}
\toprule
\textbf{Layer} & \textbf{Logical role} & \textbf{Purpose} \\
\midrule
Instruction
& Identity/persona
& Defines the agent's identity or persona. \\
& User model
& Stores instructions or assumptions about the user. \\
& Behavioral policy
& Defines rules on agent behavior. \\
& Capability guidance
& Specifies how and when available tools or capabilities should be used. \\
\midrule
Memory
& Durable global
& Stores information retained across tasks and sessions. \\
& Episodic
& Records information from individual interactions or sessions. \\
& Topic-scoped
& Stores persistent information associated with a specific topic or task. \\
\midrule
Configuration
& Runtime parameters
& Controls settings used during agent execution. \\
& Automation schedule
& Defines persistent scheduled or recurring agent actions. \\
\bottomrule
\end{tabularx}
\end{table}

\subsection{Self-States}
\label{sec:arch}

A self-hosted AI agent runs on a user's machine and can interact with the host operating system through capabilities such as shell command execution and file-system access. Its architecture consists of an LLM core, a tool interface exposing capabilities
such as a persistent shell and file system, and persistent resources that the
agent later consumes to determine how it acts. We call these resources
\emph{self-state}. This definition is based on function rather than filename, i.e., a resource belongs to self-state when changing it can influence the agent's
future decisions, tool use, or retained knowledge.

We organize file-backed self-state into the three top-level target classes in
Table~\ref{tab:self-state-roles}. Each class contains distinct logical roles.
For example, durable global memory and an episodic session log are both Memory
state, but need not share a path, loading rule, or natural update pattern.

A \emph{state schema} maps these logical roles to the concrete files or path
patterns used by an agent implementation and records each object's scope and
consumption path. The schema does not assume that every object in a layer has
the same mutability. Natural update behavior is measured under each workload.
The security property of interest is that a malicious update from an attacker can have
the same OS projection as a legitimate update to the same mapped
object (\S\ref{sec:formalization}).

% Cross-boundary defense is well-studied; self-state defense is the harder open problem, and the focus of this paper.

% This profile-dependence makes defense analysis inherently workload-conditioned.
% Write frequency is the property that connects this three-layer decomposition to the defense analysis in \S\ref{sec:gaps}: rarely-written layers admit prevention-by-denial, continuously-written layers do not, and workload-dependent layers belong to neither bucket cleanly.
% The formal argument and its consequences for each OS defense dimension live in \S\ref{sec:prevention-gap}.

\subsection{Threat Model}
\label{sec:ssa-def}
\paragraph{Attacker capability} The attacker has no direct access to the host or its kernel. The attacker can place arbitrary content into channels the agent reads (documents, web pages, messages, tool outputs) and use that content to compromise the agent's decision process via indirect prompt injection or related techniques. We treat the outcome of compromise as given: post-compromise, the agent executes the attacker behavior under its own OS principal and permissions. How compromise is achieved (the prompt-injection layer) has been studied separately~\cite{greshake2023,promptware2026,habler2026memorypoisoning,clawjacked2026}. This paper instead focuses on what a compromised agent can do to its own persistent state and what OS-layer defenses can do about it.

\paragraph{Trusted components}
We trust the OS kernel and the defense infrastructure, including the
monitoring, backup, and policy-enforcement components and the storage on
which they rely. These components execute under an OS principal distinct
from the agent's, and the agent has no legitimate write access to them.
This privilege separation protects the defense infrastructure from direct
tampering, but does not resolve the selective enforcement requirement: legitimate and malicious writes to self-state originate from the same agent principal and target the same authorized resources
(\S\ref{sec:prevention-gap}).

\paragraph{Out of scope} Direct host access, attacks on the LLM weights or training, such as jailbreak, model extraction, or supply-chain, as well as an attacker who simultaneously controls the agent and the defense components, are outside the threat model.

\subsection{Formalization}
\label{sec:formalization}

Figure~\ref{fig:self-state-attack} separates attacks by whether the relevant
operation crosses an enforceable boundary. A \emph{cross-boundary attack}
requests an operation that a correctly configured policy can exclude from the
agent's authorized scope, such as contacting a forbidden endpoint or modifying
a protected host process. Isolation is effective when this policy distinction
exists~\cite{isolategpt2025,progent2025,seagent2026,camel2025}. A
\emph{self-state attack}, by contrast, targets a resource that the agent may
legitimately update under the threat model above.

Let $L(R)$ and $A(R)$ denote legitimate and attacker-induced operations on the
self-state resource $R$. For a fixed operation-level observation model
$\mathcal{O}$, let $\pi_{\mathcal{O}}(x)$ denote the exact input that a policy
or detector receives for operation $x$. In the analysis below, this input is
restricted to the current operation's principal, object, operation kind, and
observable metadata. The projection is therefore relative to a mechanism. As a consequence, it is not
shorthand for all information that any OS-hosted component could obtain. Let
$U_{\mathcal{O}}$ be the set of inputs authorized by the policy under this
observation model. Figure~\ref{fig:self-state-attack} contrasts a
cross-boundary operation with $\pi_{\mathcal{O}}(a)\notin U_{\mathcal{O}}$
against the condition that creates the scoped decision overlap:
\[
  \exists a\in A(R),\;\exists \ell\in L(R):
  \pi_{\mathcal{O}}(a)=\pi_{\mathcal{O}}(\ell).
\]

\begin{definition}[Self-State Attack]
A self-state attack is an attacker-induced file-system operation against the
agent's persistent memory, instruction, or configuration state, issued under
the agent's legitimate authorization.
\end{definition}

We call an instance \emph{projection-overlapping} for $\mathcal{O}$ when it
satisfies the equality above with at least one legitimate update on the same
resource. The attack class is defined by its target and authorization; the
prevention and detection analysis concerns only its projection-overlapping
instances. We do not claim that every self-state attack overlaps a legitimate
update, or that an OS-hosted component may never inspect content.

% A comparison of this threat model with former ones is detailed in the Appendix~\ref{app:prior-threat-models}.

\subsection{The Attack Space}
\label{sec:attack-space}

To analyze OS defenses against self-state attacks, we organize the attack
space from the OS layer's vantage point. For the auxiliary event-level models,
an observed event $e$ is represented by the content-agnostic feature tuple
\[
\langle f,o,\delta_s,\Delta m,\Delta t\rangle,
\]
where \(f\in\mathcal{F}\), and \(\mathcal{F}\) is the set of
workspace-relative paths to monitored files. The variable \(o\) is the
discrete virtual-file-system (VFS) event kind, drawn from
\[
\begin{aligned}
o\in\{&\texttt{write},\texttt{create},\texttt{unlink},\\
        &\texttt{rename},\texttt{attribute-change}\}.
\end{aligned}
\]
The term \(\delta_s\in\mathbb{Z}\) is the signed change in file size, in
bytes. \(\Delta m=(m_{\mathrm{before}},m_{\mathrm{after}})\) records the
POSIX permission modes before and after the event, each represented as
an octal permission-bit mask from \texttt{0000} to \texttt{7777}.
\(\Delta t\in\mathbb{R}_{\geq 0}\cup\{\bot\}\) is the elapsed time, in
seconds, since the preceding event with the same \((f,o)\) key, or
\(\bot\) when no preceding event exists. These fields are obtained from
the kernel event notification together with the pre-event and post-event
metadata maintained by the collector.

The auxiliary pooled model B1 and profile-conditioned model B2 use \(f\) and
\(o\) to select the per-\((\textit{file},\textit{op\_type})\) baseline and score
\((\delta_s,\Delta t)\); the \(\Delta t\) term is omitted when it is
undefined. Permission-mode changes identify M4 attacks and are monitored
separately from payload mutation. Snapshot integrity additionally compares
file size and content hashes. This tuple is one detector-specific
instantiation of $\pi_{\mathcal{O}}$ rather than a universal OS feature vector.
Prevention, snapshot integrity, Falco, sequence novelty, and provenance each receive the
different inputs specified in Section~\ref{sec:exp-designs}. We choose four
attack-space axes that project onto one or more components of the event tuple
(Table~\ref{tab:axis-vfs}).

\begin{table}[t]
\caption{The four axes and their candidate OS-visible projections.}
\label{tab:axis-vfs}
\centering
\footnotesize
\setlength{\tabcolsep}{2pt}
\begin{tabularx}{\columnwidth}{@{}
  >{\raggedright\arraybackslash}p{.22\columnwidth}
  >{\raggedright\arraybackslash}p{.26\columnwidth}
  >{\raggedright\arraybackslash}X
  >{\raggedright\arraybackslash}p{.2\columnwidth}@{}}
\toprule
\textbf{Axis} & \textbf{Controls} & \textbf{VFS projection} & \textbf{Monitor field} \\
\midrule
D1 Target       & state class and role & resolved object path & path/fd resolution \\
D2 Mechanism    & how state diverges & event + $\operatorname{sign}(\delta_s)$ & \texttt{IN\_*} mask \\
D3 Granularity  & size of change  & $|\delta_s|$       & \texttt{size\_delta} \\
D4 Temporal     & injection time  & inter-arrival     & $\Delta t$ \\
\bottomrule
\end{tabularx}
\end{table}

\textbf{D1 (Target)} identifies the top-level state class being corrupted:
Instruction, Memory, or Configuration. A logical role from
Table~\ref{tab:self-state-roles} identifies the concrete instantiation within
that class, such as episodic rather than durable global memory. Roles preserve heterogeneity within the layer without creating additional cells in the primary
Target $\times$ Mechanism $\times$ Granularity matrix.

\textbf{D2 (Mechanism)} captures the concrete file-system operation
through which the attack is applied to the self-state. We consider four
mechanisms: \emph{M1 Modify}, which overwrites existing content in place;
\emph{M2 Add}, which appends new content or inserts it into an existing
file; \emph{M3 Delete}, which removes content by unlinking files or
truncating them; and \emph{M4 Deny}, which changes permission bits so
that subsequent reads fail. Unlike the first three mechanisms,
\emph{M4 Deny} does not directly alter file contents. However, when
loaders fail softly, an unreadable state may be treated as missing,
defaulted, or partially loaded, producing decision-level effects
comparable to content modification or insertion.

\textbf{D3 (Granularity)} describes the size of the concrete edit applied
to the self-state. We distinguish four levels: \emph{G1 whole-file},
where the attack replaces or removes an entire file; \emph{G2
large-delta}, where the edit changes tens to hundreds of bytes;
\emph{G3 small-delta}, where the edit is roughly line-scale; and
\emph{G4 minimal-delta}, where the edit changes at most a few bytes
(\(\leq 4\)~B), such as flipping a digit or toggling a Boolean value.
Granularity is the axis over which the attacker has the greatest freedom:
the same semantic manipulation may often be realized through edits of
very different sizes. Minimal-delta attacks are especially relevant to
detection because their size features may overlap the agent's legitimate write
distribution.

\textbf{D4 (Temporal)} describes how attack events are scheduled over
time. We distinguish four timing patterns: \emph{T1 single-shot}, where
a single attack event occurs within an otherwise normal session, which
is our primary configuration. \emph{T2 burst-aligned}, where the attacker
injects writes during periods of high agent activity. \emph{T3
quiet-gap}, where malicious writes are placed during periods of low agent
activity. And finally \emph{T4 slow-drip}, where the attack is distributed
across many small writes over a long time window. These patterns define
scheduling choices and do not assume that any individual write evades a
particular detector.

Throughout the paper, axis values are labeled by the initial of their axis:
mechanisms M1 to M4, granularities G1 to G4, and temporal patterns T1 to T4.
Targets are named directly (Memory, Instruction, Configuration). The structural matrix fixes the temporal axis to the primary T1
single-shot setting. A matrix cell is therefore identified by
Target-Mechanism-Granularity, as in Mem-M3-G1 for a whole-file delete of
\texttt{MEMORY.md}. T2 through T4 remain dimensions of the broader taxonomy
and are not claimed as empirical matrix coverage in this evaluation.

\subsection{Attack Space from Threat Catalogs}
\label{sec:attack-instantiation}

Our four-axis space is grounded in existing catalogs of AI-agent attacks and system-level adversarial operations, but specializes them to the self-state setting. We use three complementary anchors. MITRE ATLAS~\cite{mitreatlas2025} provides AI-specific anchors for the affected self-state target: memory-state cells map to \textsc{aml.t0080.000} (AI Agent Context Poisoning: Memory), instruction-state cells map to \textsc{aml.t0080} (AI Agent Context Poisoning), and configuration-state cells map to \textsc{aml.t0081} (Modify AI Agent Configuration). MITRE ATT\&CK for Enterprise~\cite{MITRE2024} provides OS anchors for the concrete file-system operation, such as T1565.001 Stored Data Manipulation, T1070.004 File Deletion, T1222 File and Directory Permissions Modification, and T1562 Impair Defenses. OWASP Agentic AI~\cite{owaspagenticthreats2025} provides the closest agent-level consequence categories, including Memory Poisoning, Identity Spoofing, Intent Breaking, Tool Misuse, and Privilege Compromise. We omit the OWASP numeric prefixes here to avoid collision with our T1--T4 temporal notation. The full per-cell mapping is given in Appendix~\ref{app:catalog-mapping}.

This mapping makes the space a self-state-specific decomposition of established
attack patterns rather than a detached design space. The catalog entries are
not themselves labeled self-state attacks, but describe broader forms of
memory poisoning, context poisoning, configuration modification, deletion, or
permission change. Granularity and Temporal refine these catalog concepts at
the OS observation boundary, where the same high-level attack may be realized
as a whole-file rewrite, a line insertion, a few-byte toggle, or a sequence of
small writes.

%% ===========================================================
%% SECTION 3: OS Defense Gaps
%% ===========================================================

\section{OS Defense Capabilities and Decision Requirements}
\label{sec:gaps}

Self-state defense does not begin from an absence of OS capability. The OS can
enforce access boundaries, record how state changes, and restore protected
prior versions. The open question is whether those capabilities provide the
decisions required when the protected object is also part of the agent's
authorized update surface. Table~\ref{tab:three-gaps} maps each capability to
its self-state-specific decision requirement and to the evaluation that tests
it in this paper.

\begin{table}[t]
\caption{OS capabilities, self-state-specific decisions, and corresponding
evaluations.}
\label{tab:three-gaps}
\centering
\footnotesize
\setlength{\tabcolsep}{2pt}
\renewcommand{\arraystretch}{1.08}
\begin{tabularx}{.98\columnwidth}{@{}
    >{\raggedright\arraybackslash}p{.16\columnwidth}
    >{\raggedright\arraybackslash}p{.22\columnwidth}
    >{\raggedright\arraybackslash}X
    >{\raggedright\arraybackslash}p{.21\columnwidth}@{}}
\toprule
\textbf{Dimension} &
\textbf{OS capability} &
\textbf{Required decision} &
\textbf{Evaluation} \\
\midrule
Prevention & Enforce write boundaries &
Which authorized updates may proceed? & Attack vs. legitimate blocking \\
Detection & Record mutation, object, and lineage &
Is the observed activity attack-specific? & FPR, landed-run detection, provenance \\
Recovery & Protect and restore prior state &
Which point should be restored? & Availability, integrity, rollback loss \\
\bottomrule
\end{tabularx}
\end{table}

\subsection{Prevention: Enforcement versus Authorized Updates}
\label{sec:prevention-gap}

OS access control can reliably prevent an operation when the policy mediates
every authority path that can modify protected state. A file-mode rule alone
may leave namespace operations available through a writable parent directory.
Relevant mechanisms include discretionary access control (DAC), POSIX
access-control lists, mandatory access control (MAC) systems such as SELinux
and AppArmor, extended Berkeley Packet Filter (eBPF) Linux Security Module
(LSM) hooks, inode immutability, and sandboxing interfaces such as
Landlock~\cite{landlock2022}. These mechanisms are particularly effective when
malicious and legitimate operations use different principals, objects, or
operation types.

Self-state introduces a selective-enforcement requirement. An agent that may
legitimately update a state object already holds a write path to that object,
and a malicious attacker update can reuse the same path. For an overlapping
pair with $\pi_{\mathcal{O}}(a)=\pi_{\mathcal{O}}(\ell)$, a policy restricted
to that current operation input assigns the pair the same decision. This is a
scoped consequence of the policy interface, not a claim about every OS-hosted
defense. A system can separate the pair by introducing a new boundary, such as
a trusted update broker, an explicit update phase, or finer state-specific
authority.

The corresponding evaluation therefore separates operation coverage from
selectivity. We first ask which content, namespace, and metadata operations a
control mediates. For an operation that is blocked, we then ask whether the
same policy preserves a corresponding authorized update. Keeping these
outcomes separate distinguishes an incomplete barrier from a complete but
collateral one.

\subsection{Detection: Observability versus Attack Specificity}
\label{sec:detection-gap}

Kernel observability can retain substantial evidence about a self-state
change. File notification, audit, syscall tracing, and provenance
reconstruction can identify the principal, process lineage, state object,
operation, outcome, and preceding host activity. This evidence supports
accountability and supplies grounded context for a defense decision.

The additional requirement is attack specificity. If an attacker-induced
operation and an authorized update have the same input under a detector's
observation model, a detector restricted to that input assigns them the same
score. This consequence is detector-relative: it does not say that all OS
evidence is identical, that later behavior cannot add information, or that an
OS defense cannot introduce a richer decision interface. It says that a
mutation alert, lineage edge, or syscall novelty score is not automatically an
attack verdict merely because it is OS-grounded.

Our evaluation therefore separates three questions: whether the change is
observed, whether the actor and object are attributable, and whether the
resulting decision is specific to attack rather than natural self-update.
Natural-write controls are essential for the last question as clean-no-write
controls alone measure background novelty rather than attack specificity.

\label{sec:anomaly-detection}
Workload conditioning is one possible way to enrich this decision with host
context. Different workloads can update state objects at different rates and
with different size or timing distributions. Conditioning may expose this
structure, but it may also fragment the clean corpus and raise false positives.
We therefore evaluate pooled and profile-conditioned baselines as an empirical
hypothesis, not as an assumed solution.

\subsection{Recovery: Restoration versus Recovery-Point Selection}
\label{sec:recovery-gap}

OS isolation and storage mechanisms can keep prior state versions outside the
agent's write authority and restore their bytes after corruption. This gives
recovery two concrete prerequisites. First, the recovery repository must remain
available even when the compromised agent attempts to destroy it. Second, a
trusted trigger, which may come from a detector, an operator, or another
protected component, must identify a recovery point.

Restoration success does not by itself establish that the selected point is
operationally desirable. A recent point may retain corruption, whereas an
older point may discard legitimate post-snapshot work. Our experiments measure
repository availability, byte and functional restoration, and observed
rollback loss for a known recovery point. Selective repair, corruption-onset
estimation, and consistent recovery across multiple state objects remain
outside the present evaluation.

\subsection{Relationship between Defense Dimensions}
\label{sec:gap-cascade}

The dimensions act at different stages of a state change but do not form a
strict pipeline. Prevention decides which writes reach commit. Detection
interprets committed activity and may supply a recovery trigger, but an
operator or another trusted component can also invoke recovery. Recovery, in turn, relies on protected prior versions that remain inaccessible to the agent. Therefore, its effectiveness does not depend on prevention or detection having succeeded. A write
barrier can remove required functionality, a detector can react to benign
self-update, and a successful restore can discard later legitimate work.

\begin{figure}[t]
  \centering
  \includegraphics[width=\columnwidth]{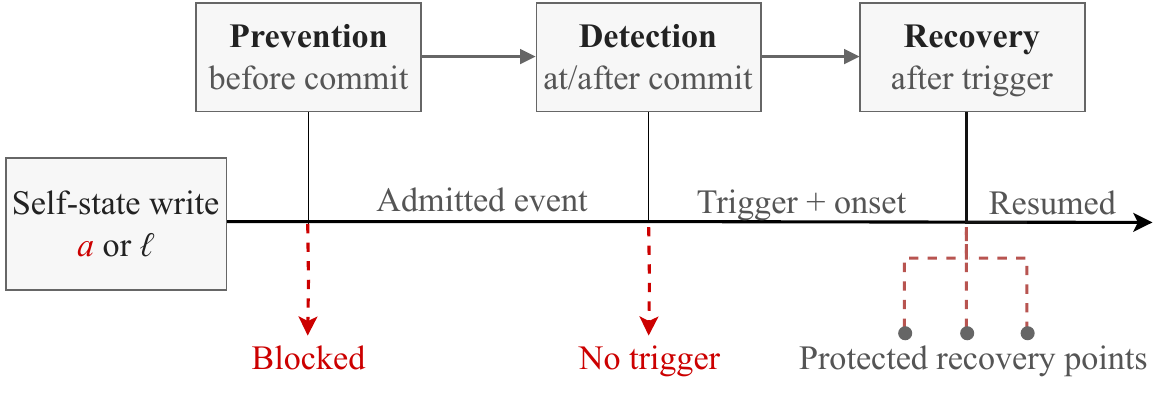}
  \Description{A horizontal OS event stream begins with either an
  attacker-induced or legitimate self-state write. Prevention gates the write
  before commit, and detection evaluates admitted activity at or after commit.
  Recovery follows a trusted trigger that may come from detection or another
  protected source and selects among recovery points kept outside the agent's
  write authority.}
  \caption{Prevention, detection, and recovery address different decisions on
  the OS-visible event stream. Prevention shapes admitted writes, detection
  assesses committed activity, and recovery follows a trusted trigger while
  relying on independently protected prior state.}
  \label{fig:defense-coupling}
\end{figure}

Figure~\ref{fig:defense-coupling} summarizes these relationships on a single
write. Each dimension consumes different evidence and leaves a different
decision unresolved. Synchronous interception can change when a decision is
enforced, finer-grained authority can create new update boundaries, richer
host context can inform attack specificity, and protected versioning can reduce
recovery cost. These are design opportunities for self-state-aware OS defenses,
not capabilities supplied automatically by generic file and syscall baselines.

%% ===========================================================
%% SECTION 4: Experimental Methodology
%% ===========================================================

\section{Experimental Methodology}
\label{sec:experiments}

Our evaluation asks three questions derived from the decision requirements in
Section~\ref{sec:gaps}. First, which self-state mutation paths do OS controls
mediate, and what functionality cost accompanies a complete barrier? Second,
how do host-level detectors trade attack coverage against specificity on
legitimate self-state activity, and do paired provenance features differ between
model-driven attack landings and task-matched clean branches? Third, can
protected recovery storage survive same-agent tampering and restore a known
point, and how much legitimate state does that rollback discard?
Before answering these questions, we validate that the observation substrate
can witness the operations on which the evaluation depends. This validation is
a measurement prerequisite, not an additional defense claim.

\begin{figure*}[t]
  \centering
  \includegraphics[width=\textwidth]{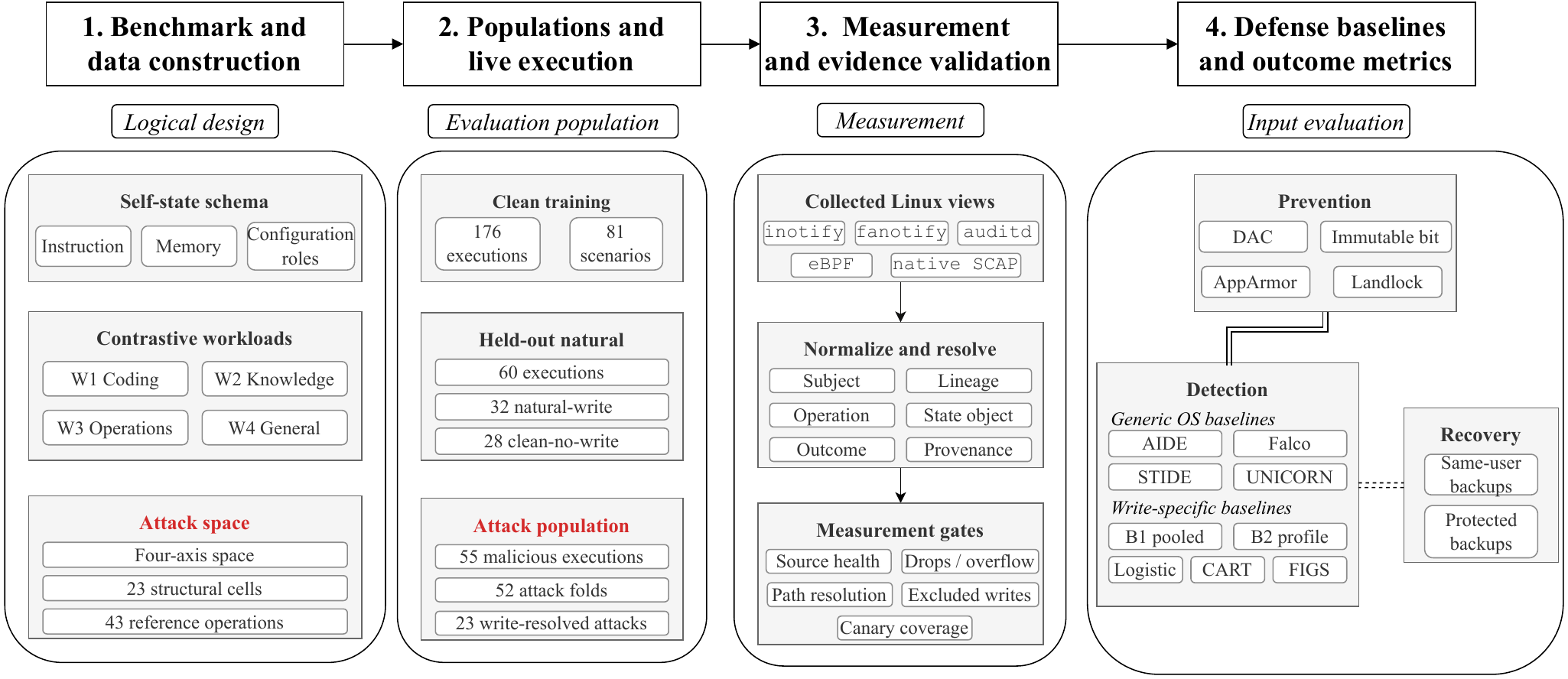}
  \Description{Evaluation pipeline in which logical self-state
  roles are instantiated as an attack catalog and executable
  operations. Clean and attack executions pass quality gates before
  entering defense evaluations, while a separate operation
  observability validation suite checks whether the collection substrate
  witnesses the required operation classes.}
  \caption{Benchmark and evaluation pipeline. Catalog
  entries, attack executions, and validation cases are
   evidence populations while model-driven landings form a subset of the
  attack executions. Measurement quality gates are evaluated before
  defense outcome metrics.}
  \label{fig:method-pipeline}
\end{figure*}

Figure~\ref{fig:method-pipeline} summarizes the evaluation pipeline and
separates benchmark construction, population selection, measurement
sufficiency, and defense effectiveness. This separation prevents structural
catalog size or sensor breadth from being mistaken for a defense result.

\subsection{Reference Adapter and Contrastive Workloads}
\label{sec:platform}

The benchmark addresses self-state through the logical schema introduced in
\S\ref{sec:arch}. Attacks and measurements retain both the logical state role
and its concrete path binding. Selecting another file-backed agent therefore
requires a new state manifest and execution backend, rather than a new
definition of the attack space or OS observation model.

For the experiments in this paper, the reference backend is
\texttt{openclaw-core}, a lightweight implementation of the agent loop, tool
dispatch, and file-backed persistent-state loading paths needed for our
OpenClaw-oriented host-level experiments. The accompanying manifest maps its
concrete files to the logical roles in Table~\ref{tab:self-state-roles}. Each
run starts from a recorded workspace seed and executes under an unprivileged
agent principal, producing ordinary Linux files and processes rather than
simulator-only events.

We use four workload profiles to create contrasting legitimate activity
regimes: coding (W1), knowledge work (W2), operations (W3), and general
assistance (W4). W1 is drawn from the Aider polyglot benchmark~\cite{aider},
W2 from FRAMES~\cite{frames2024}, and W3--W4 are version-controlled authored
tasks. The profiles are not claimed to estimate a population distribution of
all agent users. Their purpose is to exercise different combinations of code
editing, information synthesis, self-configuration, and everyday file
operations.

\begin{table}[t]
\caption{Clean-corpus composition by workload profile. The held-out set has
five scenarios and three executions per scenario in each profile.}
\label{tab:profiles}
\centering
\footnotesize
\setlength{\tabcolsep}{3pt}
\begin{tabular}{@{}llrrr@{}}
\toprule
\textbf{Profile} & \textbf{Task regime} & \textbf{Train exec.} &
\textbf{Test scen.} & \textbf{Test exec.} \\
\midrule
W1 & Coding assistant   & 45 & 5 & 15 \\
W2 & Knowledge worker   & 45 & 5 & 15 \\
W3 & Operations agent   & 51 & 5 & 15 \\
W4 & General assistant & 35 & 5 & 15 \\
\midrule
Total & & 176 & 20 & 60 \\
\bottomrule
\end{tabular}
\end{table}

\subsection{Evaluation Populations}
\label{sec:populations}

We construct clean and attack evidence independently. The clean training
population contains 176 executions from 81 scenarios. The held-out natural
population contains 60 executions from 20 scenarios, with
three replicates per scenario: 32 natural-write and 28 clean-no-write
executions. Training executions fit learned baselines, while the held-out
corpus is used only for false-positive evaluation. Point estimates use
executions, while uncertainty resamples scenarios with all replicates retained.

The attack population contains 55 adjudicated malicious executions grouped
into 52 attack folds. 35 involve model-driven writes in which
attacker-induced content persisted, while 20 instantiate metadata or namespace
mutations directly under the compromised agent principal. Under our threat
model, which conditions on successful agent compromise, both groups are attack
executions for evaluating post-compromise OS defenses. The distinction records
whether an execution additionally demonstrates end-to-end model inducibility, and does not affect inclusion in the attack population. Write-specific methods
are defined on the 23 attacks with resolvable self-state writes and valid size
features. Attacks outside that mechanism definition are reported as uncovered
rather than negative predictions.

For a secondary provenance analysis, we compare 21 model-driven operational
landings with 21 task-matched clean branches. These controls ask whether actor,
object, and carrier relations differ between an induced self-state write and
its corresponding clean activity. They are not used to train the primary
detector comparison or to estimate natural-workload false positives.

\subsection{Attack Operationalization}
\label{sec:attack-suite}

Fixing the Temporal axis to T1, the valid Target $\times$ Mechanism
$\times$ Granularity projection of Section~\ref{sec:attack-space} contains the
23 structural cells in Table~\ref{tab:attack-matrix}. For the reference
adapter, the executable catalog expands these cells to 43 parameterized
operations, each carrying a logical state role and a concrete path binding.
The matrix is a coverage scaffold: it defines what an OS defense should be
asked to observe or control. Catalog entries and attack executions are
different counting units: the former describe parameterized operation
bindings, whereas the latter are run-level attack records.

\begin{table}[t]
\caption{The 23-cell structural attack matrix at T1. Entries list cataloged
granularities. n/a denotes a combination excluded by construction.}
\label{tab:attack-matrix}
\centering
\footnotesize
\setlength{\tabcolsep}{2pt}
\begin{tabularx}{\columnwidth}{@{}
  >{\raggedright\arraybackslash}p{.22\columnwidth}
  *{4}{>{\centering\arraybackslash}X}@{}}
\toprule
\textbf{Target} & \textbf{M1 Modify} & \textbf{M2 Add} &
\textbf{M3 Delete} & \textbf{M4 Deny} \\
\midrule
Memory      & G1--G4 & G2--G4 & G1 & G1 \\
Instruction & G1--G4 & G2--G4 & G1 & G1 \\
Configuration & G1--G4 & n/a & n/a & G1 \\
\bottomrule
\end{tabularx}
\end{table}

The 55 attack executions are divided into two groups. In 35 model-driven attacks, malicious content is introduced through a poisoned user message, an ingested carrier file, or tool-delivered content. The agent then runs normally, and a before/after marker test determines whether the attack produced a persistent self-state modification. The remaining 20 attacks directly execute metadata or namespace mutations under the compromised agent principal. These cases test whether OS defenses can handle the resulting malicious operation, rather than whether that operation can be elicited through prompt injection. Both groups are included in the OS-level attack evaluation.

Catalog entries without an admitted attack execution remain
catalog-only and are not counted as empirical attack coverage. Direct
instantiation should also not be confused with the operation observability
validation in Section~\ref{sec:trace-collection}: the former contributes
adjudicated attack executions to the attack population, whereas the
latter is a separate measurement-only population used to validate collection
coverage.

\subsection{Host Telemetry and Evidence Validation}
\label{sec:trace-collection}

No single host source supplies all identities needed by the baselines. We
therefore collect five host views for each run: \texttt{inotify},
\texttt{fanotify}, \texttt{auditd}, extended Berkeley Packet Filter (eBPF)
telemetry, and native Sysdig capture (SCAP).

\begin{table}[t]
\caption{Collected sources and their role. Collection breadth is distinct from
detector consumption: no detector is described as consuming all five sources.}
\label{tab:five-source}
\centering
\footnotesize
\setlength{\tabcolsep}{4pt}
\begin{tabularx}{\columnwidth}{@{}lX@{}}
\toprule
\textbf{Source} & \textbf{Role in the measurement} \\
\midrule
\texttt{inotify} & Independent file-event chronology and overflow check \\
\texttt{fanotify} & File-operation evidence used by normalization \\
\texttt{auditd} & Syscall, credential, and path records; adjudication channel \\
eBPF & Process/syscall identity and timing correlation \\
SCAP & Native Sysdig capture for libsinsp file-descriptor reconstruction and Falco replay \\
\bottomrule
\end{tabularx}
\end{table}

The normalizer combines audit, eBPF, fanotify, and Sysdig-capture evidence into
a syscall stream and provenance graph. \texttt{inotify} remains an independent
file-event view rather than being silently counted as another graph edge.
Detector inputs are then exported in detector-native form: the Advanced
Intrusion Detection Environment (AIDE) receives before/after state snapshots,
Falco replays the native SCAP capture, sequence time-delay embedding (STIDE)
receives normalized per-process syscall sequences, and provenance analyses
receive normalized nodes and edges.

A run is admitted only if source-health checks pass, the effective
file-descriptor-to-path resolution rate is at least 0.95, no self-state write is
excluded by the normalizer, and no collector reports an invalid status. All 176
training and 60 held-out clean executions have non-empty five-source inputs,
zero recorded drop/overflow, no excluded writes, and an effective
path-resolution rate of 1.0. Training and held-out scenarios are disjoint.

We separately validate operation observability using a 16-case validation
suite: write, rename, unlink, and chmod crossed with memory-root, memory-log,
instruction, and configuration targets. Every case must be witnessed in the
normalized graph. The validation suite is not part of the admitted detector
corpus, so the 0.95 corpus threshold is neither relaxed nor reused for it. One
of nine file-descriptor-bearing graph operands is unresolved: an
\texttt{auditctl} netlink socket rather than a self-state mutation target. All
16 operation-by-target-role cases nevertheless retain a graph witness. We
therefore use this validation only for the per-case witness endpoint, not for a
corpus-level path-resolution claim. The result establishes that the agent
principal can issue each tested operation and that the collection path records
it. The validation suite contributes no attack observations and does not
affect the 55-execution attack denominator.

Ground truth combines immutable run manifests, before/after snapshots, the
logical-object-to-path bindings, and separately retained raw audit records.
Clean runs with no mapped self-state change are retained rather than selected
away.

Taken together, these controls establish the scope within which the evidence is
sufficient. Collector failures are ruled out through source-health and drop
checks, while file-descriptor-to-path resolution accounts for every relevant
write at the object level. The observability validation confirms that the
measurement substrate captures the tested operations under each state role.
Detector evaluation remains isolated from training through pre-specified,
scenario splits, and matched clean branches show whether provenance
features are specific to model-driven operational landings. Maliciousness is
determined separately through the pre-specified adjudication protocol for all
55 attack executions.

\subsection{Defense Evaluations}
\label{sec:exp-designs}

\subsubsection{Prevention}
\label{sec:exp1-design}

We evaluate whether a file-level policy provides both operation coverage and
update selectivity on the durable-global memory binding, \texttt{MEMORY.md}.
The paired direct-write probe compares an attack-designated marker write with
an operator-approved direct write by the same unprivileged runner. It covers
root-owned read-only DAC, inode immutability, AppArmor, and
Landlock~\cite{landlock2022}, and determines success from the post-state bytes
and the kernel error code (\texttt{errno}).

A supplemental replay tests five configurations, adding self-owned read-only
DAC, against direct overwrite, append, truncate, rename-over, unlink, and
permission-reset followed by overwrite. A trusted supervisor or launcher
installs each policy before the agent-principal probe. The replay verifies the
requested post-state rather than inferring success from policy installation.
It measures operation coverage while the paired direct-write probe separately
measures whether blocking also rejects a corresponding authorized update.
Exact platform and post-state checks appear in
Appendix~\ref{app:prevention-replay}.

\subsubsection{Detection Baselines}
\label{sec:detection-design}

The detector comparison spans four common forms of host evidence. AIDE uses
before/after file integrity, Falco applies pre-specified runtime rules to SCAP
events, STIDE detects unseen six-call syscall sequences in core processes using workload-specific normal profiles, and UNICORN~\cite{han2020unicorn} models role-typed provenance
graphs. We also evaluate pooled and workload-conditioned statistical anomaly
baselines over write size and timing, together with three small supervised
models. Configurations, features, and decision thresholds are selected without
held-out natural executions.

The unit of decision is one execution. We report attack TPR together with the
attack population on which a method is defined, and natural-workload FPR on the
separate held-out corpus. Failed parsing or analyzer execution is
data-insufficient rather than a negative prediction. Natural-write and
clean-no-write controls are also separated because a mutation detector can
appear sensitive to attacks simply by flagging any self-state update. Complete
decision and evaluability rules appear in Appendix~\ref{app:baseline-rules}.

\subsubsection{Provenance Attribution}

For each of the 21 model-driven operational landings and its task-matched clean
branch, we ask whether the OS can resolve the destination object, attribute the
writer to the agent process tree, and connect an ingested filesystem carrier
to that writer. User messages and content delivered by the network have no filesystem
carrier read and are therefore not evaluable on the last property. The matched
comparison tests whether provenance distinguishes an induced write, rather than
only naming its actor and object.

\subsubsection{Statistical Anomaly and Supervised Baselines}
\label{sec:exp2-temporal-design}

For each resolved self-state write event \(e\), we form a key \(k\) from its
state-object bucket and operation type and let
\(q(e)=\log\max\{\Delta t(e),1\,\mathrm{ms}\}\). We then compute
\begin{equation}
\mathrm{score}(e;k)=\max\!\left\{
  \frac{|\delta_s(e)-\mu_s(k)|}{\sigma_s^{\mathrm{eff}}(k)},
  \frac{|q(e)-\mu_t(k)|}{\sigma_t(k)}
\right\}.
\label{eq:anomaly-score}
\end{equation}
Here \(\mu_s(k),\sigma_s(k)\) and \(\mu_t(k),\sigma_t(k)\) are the
clean-training means and sample standard deviations of signed size change and
log inter-write time, respectively, and
\(\sigma_s^{\mathrm{eff}}(k)=\max\{\sigma_s(k),0.1|\mu_s(k)|,1\,\mathrm{B}\}\).
The timing term is omitted when a preceding same-key write or an adequate
clean timing baseline is unavailable. B1 fits pooled clean data, B2 fits each
workload profile separately, and both use the pre-specified threshold
\(\mathrm{score}(e;k)\geq2\). Appendix~\ref{app:baseline-rules} specifies
the remaining boundary cases.

The supervised comparison uses \(L_1\)-regularized logistic regression, a
classification and regression tree (CART), and Fast Interpretable Greedy-Tree
Sums (FIGS) over the same write-resolved population and related host features.
Attack predictions use five scenario-grouped outer folds; each model trains on
the remaining attack folds and all clean training executions, with its
threshold selected by grouped inner cross-validation. Natural FPR is measured
by refitting on the full training population and applying the model once to
the independent natural corpus. Neither held-out natural executions nor
matched twins enter feature or threshold selection.
Appendix~\ref{app:baseline-rules} records the full feature set and
pre-specified evaluation protocol.

\subsubsection{Recovery}
\label{sec:exp3-design}

Recovery begins from a post-session workspace, creates a snapshot, applies one
legitimate post-snapshot update, executes a canonical corruption, and invokes
restore. The repository experiment crosses protected-principal versus
same-user backup storage with ordinary corruption versus an attack that also
destroys reachable backups. A protected supplement repeats restoration for
eight attacks spanning the state roles and operation families.

We report backup availability, byte-identical restoration, an independent
health check, rollback loss, and restore latency separately. Because the
protocol fixes one post-snapshot update, its one-path rollback loss is a design
constant rather than an estimate of natural workload cost. We therefore
characterize per-session rollback exposure using all 236 clean executions from
the training and held-out natural corpora. This combined corpus is used only
for descriptive recovery-cost characterization. We do not extrapolate these
session summaries across longer backup intervals because repeated changes to
one object collapse under the recovery metric.

\subsection{Experimental Scope}
\label{sec:vci-formalism}

The design supports different claims at different scopes. Detection point
estimates use executions, but uncertainty resamples the 20 clean scenarios or
52 attack folds with all correlated executions retained. The 60 held-out
natural executions are therefore not 60 independent units, and profile-level
rates based on five scenarios remain descriptive. Boundary-cell and complete
interval procedures appear in Appendix~\ref{app:baseline-rules}.

The 23-cell matrix establishes structural catalog coverage, while the 55
attack executions define the realized OS-level attack population under our
post-compromise threat model. Within that population, the distinction between
35 model-driven and 20 directly instantiated cases limits only claims about
end-to-end model inducibility. Both groups contribute to the evaluation of
post-compromise OS defenses. Conditional rates for write-specific methods are
reported only on their defined population and always alongside coverage. The
matched provenance pairs support attribution contrasts, rather than detector training
or the primary natural FPR.

Prevention and recovery results apply to the tested mechanisms, object, and
platforms rather than all possible policies or backup designs. The schema
interface permits rebinding to other file-backed layouts, but all live results
use the reference backend described in Section~\ref{sec:platform}. The four
workloads are contrastive activity regimes, not a probability sample of
deployments.

%% SECTION 5: Results
%% ===========================================================
\section{Results}
\label{sec:results}

Before comparing defenses, we verify that the measurement pipeline reliably captures the events required by the evaluation. All clean training and held-out natural executions pass the five-source admission criteria, with no recorded drops, overflows, excluded self-state writes, or unresolved relevant paths. Training and held-out scenarios are disjoint. The canary tests further confirm that all 16 tested combinations are observed in the normalized provenance graph. These results establish that the measurement pipeline covers the tested operations, but do not constitute additional attack executions.
\subsection{Prevention}
\label{sec:exp1-results}

A policy that blocks direct writes may still leave other ways to replace or
remove the same state object. We therefore evaluate operation coverage before
asking whether a complete barrier preserves authorized updates.

\begin{table}[t]
\caption{Six-operation prevention replay on \texttt{MEMORY.md} under the agent
principal. The bypass column lists operations whose requested post-state landed;
parenthesized symbols are kernel errors. ``All six'' covers direct write,
append, truncate, rename-over, unlink, and \texttt{chmod/write}. The paired
direct-write probe separately tests authorized-update selectivity.}
\label{tab:prevention-results}
\centering
\footnotesize
\setlength{\tabcolsep}{3pt}
\begin{tabularx}{\columnwidth}{@{}
  >{\raggedright\arraybackslash}p{.25\columnwidth}
  >{\raggedright\arraybackslash}X
  >{\raggedright\arraybackslash}p{.24\columnwidth}@{}}
\toprule
\textbf{Control} & \textbf{Blocked operations} & \textbf{Bypassed operations} \\
\midrule
Self-owned \texttt{0444} & Direct, append, truncate (\texttt{EACCES}) &
Rename-over, unlink, \texttt{chmod/write} \\
Root-owned \texttt{0444} & Direct, append, truncate (\texttt{EACCES});
\texttt{chmod/write} (\texttt{EPERM}) & Rename-over, unlink \\
\texttt{chattr +i} & All six (\texttt{EPERM}) & None \\
AppArmor & All six (\texttt{EACCES}) & None \\
Landlock & All six (\texttt{EACCES}) & None \\
\bottomrule
\end{tabularx}
\end{table}

Table~\ref{tab:prevention-results} validates the expected operation-level differences among the tested configurations and shows why direct-write tests are insufficient. Removing the file's write bit blocks overwrite, append, and truncate, but parent-directory authority still permits replacement and removal. Root ownership additionally prevents the agent from restoring write permission, yet rename-over and unlink remain available.

Inode immutability, AppArmor, and Landlock block all six replayed operations,
showing that the OS can enforce a complete barrier over this test set. The
remaining limitation is selectivity. The paired probe directly shows that the
same policies also reject an approved direct update, and the content-blind
policy input is identical for corresponding authorized operations. None of the
tested configurations therefore combines full operation coverage with
preservation of corresponding self-updates.

\subsection{Detection}
\label{sec:exp2-results}

Attack sensitivity is meaningful only together with operation coverage and false positives on natural agent activity. Table~\ref{tab:fpr-legitimate} therefore reports all three quantities.

\begin{table}[t]
\caption{Primary detector comparison. Attack coverage is the number of attack
executions for which a method produces a decision. TPR and FPR use the
corresponding attack and natural populations. Complete intervals and
clean-control breakdowns appear in
Table~\ref{tab:detector-details}.}
\label{tab:fpr-legitimate}
\label{tab:exp2-fim}
\centering
\footnotesize
\setlength{\tabcolsep}{3pt}
\begin{tabularx}{\columnwidth}{@{}>{\raggedright\arraybackslash}X c c c@{}}
\toprule
\textbf{Method} & \textbf{Attack coverage} & \textbf{Attack TPR} &
\textbf{Natural FPR} \\
\midrule
\multicolumn{4}{@{}l}{\textit{File, syscall, and graph baselines}} \\
AIDE & 55/55 & 52/55 & 32/60 \\
Falco & 55/55 & 43/55 & 21/60 \\
STIDE & 55/55 & 55/55 & 58/60 \\
UNICORN & 41/55 & 35/41 & 35/47 \\
\midrule
\multicolumn{4}{@{}l}{\textit{Write-specific feature baselines}} \\
B1 pooled & 23/55 & 19/23 & 10/60 \\
B2 per-profile & 23/55 & 20/23 & 7/60 \\
\(L_1\) logistic regression & 23/55 & 15/23 & 7/60 \\
CART & 23/55 & 15/23 & 3/60 \\
FIGS & 23/55 & 13/23 & 3/60 \\
\bottomrule
\end{tabularx}
\end{table}

The broad file and syscall baselines detect most or all attack operations, but
their alerts are not specific to attacks. AIDE and Falco fire only when the
natural execution modifies self-state, whereas STIDE also flags 26 of 28
clean-no-write executions. Thus, the former primarily act as mutation
indicators, while the latter additionally reacts to benign syscall novelty.

The graph baseline is only partially evaluable and flags both natural-write and
clean-no-write executions at its pre-specified threshold. Its result supports
the presence of provenance signal, but not a low false-positive decision
boundary. Graphs that cannot be evaluated are treated as insufficient data rather than as negative predictions, which would artificially improve specificity.

Write-specific features reduce natural false positives, but their scope is
narrower. B2 makes only a modest improvement over pooled B1, and the supervised
models further reduce natural positives only by accepting lower conditional
TPR. Because every method in this group is defined on 23 of 55 attacks, these
conditional rates do not erase the uncovered operation classes. No evaluated
detector therefore dominates on both attack coverage and natural-workload
specificity. Complete cluster intervals and clean-control decompositions appear
in Table~\ref{tab:detector-details}. The intervals treat the 60 natural
executions as 20 scenario clusters.

\begin{table}[t]
\caption{Provenance properties on 21 operational landings and 21 matched clean
branches. Counts characterize host attribution, not semantic attack
detection.}
\label{tab:provenance-results}
\centering
\footnotesize
\setlength{\tabcolsep}{2pt}
\begin{tabularx}{\columnwidth}{@{}
  >{\raggedright\arraybackslash}X
  >{\centering\arraybackslash}p{.14\columnwidth}
  >{\centering\arraybackslash}p{.20\columnwidth}
  >{\centering\arraybackslash}p{.20\columnwidth}@{}}
\toprule
\textbf{Property} & \textbf{Landing} & \textbf{Matched clean} &
\textbf{Attack-specific?} \\
\midrule
Destination nameable & 21/21 & 21/21 & no \\
Agent principal attributable & 21/21 & 21/21 & no \\
Filesystem carrier evaluable & 13/21 & 13/21 & no \\
Chain present, if evaluable & 12/13 & 13/13 & no \\
Carrier not file-observable & 8/21 & 8/21 & no \\
\bottomrule
\end{tabularx}
\end{table}

Provenance provides reliable attribution but not attack-specific discrimination. The modified object and agent-tree writer are resolved in every landing and matched clean branch. Filesystem carriers are evaluable for 13 branches per arm, and their read-to-write chains are present at nearly the same rate. The remaining carriers enter through user or network channels and have no filesystem read for the OS to connect. Provenance can therefore answer who changed which object, but assigns the same attribution to corresponding attack and legitimate activity.

\subsection{Recovery}
\label{sec:exp3-results}

Restoration is useful only if the recovery state remains outside the
compromised agent's authority. We first test that trust boundary, then separate
successful restoration from the legitimate work it rolls back.

\begin{table}[t]
\caption{Recovery repository isolation matrix. Rollback loss is applicable
only when restore succeeds; every case contains one post-snapshot changed
path.}
\label{tab:recovery-isolation}
\centering
\footnotesize
\setlength{\tabcolsep}{2pt}
\begin{tabularx}{\columnwidth}{@{}
  >{\raggedright\arraybackslash}X
  *{4}{>{\centering\arraybackslash}p{.12\columnwidth}}@{}}
\toprule
\textbf{Repository / attack} & \textbf{Backup} & \textbf{Bytes} &
\textbf{Health} & \textbf{Loss} \\
\midrule
Protected / ordinary & yes & restored & pass & 1 \\
Same user / ordinary & yes & restored & pass & 1 \\
Protected / backup destruction & yes & restored & pass & 1 \\
Same user / backup destruction & no & failed & n/a & n/a \\
\bottomrule
\end{tabularx}
\end{table}

Both repository modes restore ordinary corruption, but only protected storage
survives an attack that also destroys agent backups. The protected
supplement confirms byte-identical and functional restoration across all
eight tested attacks. Thus, the decisive factor is whether the compromised
principal can reach the recovery state, not whether the restore command can
reproduce known-good bytes.

Every successful isolation-matrix restore removes the one legitimate update
inserted by the protocol. That one-path loss is therefore a design constant.
The independent clean-workload analysis shows why the practical exposure cannot
be summarized by that constant alone: the median session changes one
self-state object, but profile means range from 0.15 to 2.09 objects. Rollback
cost is workload-dependent even when restoration succeeds. These per-session
statistics do not determine loss over longer backup intervals, for which
repeated changes to the same object may collapse.

\subsection{Overall Defense Capabilities and Limitations}
\label{sec:synthesis}

Across the evaluated detectors, attack coverage and natural-workload specificity vary widely despite a common validated host-measurement substrate. Attack coverage ranges from 23/55 to 55/55, while natural false positives range from 3/60 to 58/60. This spread indicates that basic observability is not the main limiting factor; the remaining gap lies in how the evaluated baselines select and interpret the available host evidence.

A controlled comparison points in the same direction. B1 and B2 use the same size-and-timing features and the same frozen threshold, but conditioning on workload profile reduces natural false positives from 10/60 to 7/60 while slightly increasing conditional attack detection from 19/23 to 20/23. This improvement is modest, however, and does not remove the broader coverage-specificity tradeoff. The tradeoff also appears across mechanism families with substantially different designs, making it unlikely to be an artifact of a single detector or threshold.

The prevention and recovery results expose the same need for more selective decision structure. Inode immutability, AppArmor, and Landlock close all six tested mutation paths, but also reject the corresponding approved direct update. Protected recovery storage survives same-principal backup destruction and restores all tested corruptions, but recovery still requires selecting a trustworthy prior state and incurs workload-dependent rollback loss. Taken together, the experiments show that the OS can enforce, observe, attribute, and restore self-state changes, while the evaluated generic mechanisms do not by themselves provide a selective decision boundary for them.

\section{Discussion}
\label{sec:discussion}

Our results show that the main challenge in protecting self-state is not simply a
lack of OS capability. This section discusses what this
means for the design of self-state-aware defenses and which findings we expect
to extend beyond the evaluated Linux environment.

\paragraph{Self-state needs an update boundary.}
The prevention experiments show that strong OS enforcement is possible.
Immutable files, AppArmor, and Landlock block all six tested mutation paths.
However, a barrier that simply denies modification also prevents the agent
from performing the corresponding legitimate update. Conversely, weaker
file-level restrictions preserve more functionality but leave alternative
mutation paths available. A self-state defense therefore needs more than a
stronger deny rule. It needs a way to distinguish ordinary agent execution
from an authorized self-state update. This boundary could take the form of a
trusted update broker, an explicit update phase, or separate authority for
different self-state roles. The important property is that updating self-state
becomes a distinct security decision rather than an ordinary consequence of
the agent's filesystem permissions.

\paragraph{Host context is useful, but must be used selectively.}
The detection experiments show a similar tradeoff. Broad integrity, rule, and
syscall detectors cover many attacks, but also flag substantial natural
self-state activity. More selective statistical and supervised baselines reduce
false positives, but apply only to the subset of attacks that produce the
write events they model. At the same time, the provenance pipeline reliably
identifies the writer, process lineage, operation, and affected object. The
host therefore has useful evidence even when a generic detector cannot make
the final decision. A self-state-aware detector could combine this evidence
with information that generic baselines ignore, such as the logical role of
the state, the current agent task, or
whether the agent is in an expected update phase. Our results do not establish
which combination is sufficient, but they show why relying only on generic
file or syscall activity is unlikely to provide both broad coverage and low
false-positive rates.

\paragraph{A self-state-aware OS defense.}
Taken together, these findings suggest a simple design direction. Self-state
should be identified explicitly, its updates should cross a distinguishable
boundary, and the resulting changes should remain attributable and recoverable.
The OS already provides most of the underlying mechanisms needed for these
tasks. What is missing is the self-state-specific context that connects them.
Rather than replacing existing access control, telemetry, and recovery
mechanisms, a specialized defense could use them as enforcement and evidence
substrates while making the update decision at a boundary designed for
behavior-governing state.

\paragraph{Portability beyond the evaluated Linux stack.}
Our implementation is specific to Linux, but the main limitations we observe are
not. On macOS and Windows, self-state updates still reach persistent storage
through OS-mediated operations, and the agent may still use the same authority
to perform legitimate and attacker-induced updates. We therefore expect the
central selectivity problem to remain. The measurement pipeline itself would
need to change: mechanisms such as FSEvents and Endpoint Security on macOS, or
ETW and filesystem auditing on Windows, expose different events, identities,
and attribution guarantees compared to the Linux sources evaluated here. Similar
revalidation would be needed in containerized deployments, where overlay
filesystems can change the relationship between observed paths and underlying
objects, while namespaces and cgroups can provide additional process and
workload context. We therefore expect the qualitative enforcement-versus-
selectivity tradeoff to carry across these environments, but not the measured
coverage, false-positive rates, attribution completeness, or recovery costs.
Establishing those quantities requires platform-specific experiments.
%% SECTION 6: Related Work
%% ===========================================================
\section{Related Work}
\label{sec:related}

This section positions our work within the literature on self-state attacks. We review threat models for agentic AI, attacks against LLM-based agents and their persistent state, and defenses operating across semantic, agent-runtime, and OS layers.

\paragraph{Threat models for agentic AI}
Christodorescu~\etal~\cite{christodorescu2025} apply a systems-security lens to agentic computing, with case studies (including agents modifying their own configuration and memory) mapped to violated principles such as TCB Tamper Resistance, but stop at per-case mapping. Industry threat frameworks for agentic AI have begun naming self-state corruption as a top-level entry: MITRE ATLAS~\cite{mitreatlas2025} lists Memory Poisoning under Persistence (AML.T0080.000), OWASP Agentic Top~10~\cite{owaspagentic2026} as ASI06, and Microsoft's failure-mode taxonomy~\cite{microsoftfailuremodes2025} alongside a Memory Hardening control pillar; CSA MAESTRO~\cite{maestro2025} instead distributes self-state threats across its layered architecture without a unifying entry. These frameworks list self-state instances across life-cycle stage, layer, or harm type. Our work complements them by defining an operation-oriented dimension space and connecting its overlapping cases to prevention, detection, and recovery measurements.

\paragraph{Attacks against LLM-based agents.}
Greshake~\etal~\cite{greshake2023} demonstrated indirect prompt injection against LLM-integrated applications, and Liu~\etal~\cite{formalprompt2024} formalized and systematically benchmarked it. The Promptware Kill Chain~\cite{promptware2026} analyzes 36 studies and real-world incidents and organizes prompt-injection-driven malware delivery into a seven-stage chain whose Persistence stage covers memory and retrieval poisoning; AIShellJack~\cite{aishellljack2025} shows how agentic coding editors (Cursor, GitHub Copilot) can be hijacked into arbitrary shell commands. Our framing differs from the kill-chain view: we treat persistent corruption of agent state as a target class organized along structural dimensions rather than as a stage in an attacker's progression. A line of prior work then demonstrated specific attack instances against agent memory and knowledge bases: PoisonedRAG~\cite{poisonedrag2024} corrupts RAG retrieval via content-layer injection. AgentPoison~\cite{agentpoison2024} introduces backdoor attacks on agent long-term memory, whereas MINJA~\cite{minja2025} reduces this to query-only access. Trust-score and sanitization defenses have also been explored as semantic mitigations~\cite{memorypoisoning2026}; our paper instead studies the metadata available after a self-state operation reaches the host. Ying~\etal~\cite{openclawsecurity2026} propose a tri-layered risk taxonomy for OpenClaw~\cite{openclaw}, namely cognitive, execution, and information-system layers. ClawJacked~\cite{clawjacked2026} provides instances of these risks via a loopback exemption enabling full agent takeover. Agent-security benchmarks~\cite{agentdojo2024,injecagent2024,wasp2025,toolsword2024,agentharm2025,rjudge2024} evaluate prompt-injection risk, tool misuse, multi-step harm under jailbreak, or LLM-as-judge over interaction records, primarily at the web-navigation or tool-call surface. ASB~\cite{asb2025} adds memory through RAG-style retrieval and stored-content poisoning rather than the agent's own configuration, identity, or persona files on the host filesystem. Hardy~\cite{hardy1988} defined the original confused deputy, and ConfusedPilot~\cite{confusedpilot2024} extended the pattern to RAG. Self-state attacks are a persistent instance of this pattern in which the protected object is also part of the deputy's future decision context.

\paragraph{Defense systems.}
AgentSight~\cite{agentsight2025} uses eBPF to correlate kernel events with TLS LLM traffic and detects prompt injection from system-level signals. AgentSight provides general OS-layer observability across agent syscalls, while we focus on the self-state target and the distinction between operation attribution and semantic polarity. Other systems act at a richer boundary than our operation-level projection: IsolateGPT~\cite{isolategpt2025} and Prompt Flow Integrity~\cite{promptflowintegrity2025} use plugin and control-flow isolation; CaMeL~\cite{camel2025} restricts how untrusted retrieved data affects program flow; SEAgent~\cite{seagent2026} uses attribute-based access control and an information-flow graph; and Progent~\cite{progent2025} enforces tool-call privilege control. ToolEmu~\cite{toolemu2024} adds a pre-deployment view through LM-emulated sandboxing. Information-flow control~\cite{fides2025} likewise adds data-dependency context that lies outside our metadata-only claim. Rule-based AgentSpec~\cite{agentspec2026}, learned-policy AgentGuardian~\cite{agentguardian2026}, GuardAgent~\cite{guardagent2024}, and VeriGuard~\cite{veriguard2025} add policy or semantic context before execution. These approaches are complementary: they may separate cases that overlap under \(\pi_{\mathcal{O}}\), while our experiments measure what remains available at the host operation boundary.

%% ===========================================================
%% SECTION 7: Conclusion
%% ===========================================================
\section{Conclusion}
\label{sec:conclusion}
Self-state attacks expose a boundary condition for OS defense: an attacker can
corrupt behavior-governing state through the same principal, objects, and
operation paths that support legitimate adaptation. We asked how far
representative OS mechanisms can protect this shared write surface across
prevention, detection, and recovery. The answer is not that the OS lacks defensive value. In our experiments, it
enforces complete barriers over the tested mutation paths, reconstructs who
changed which state object, and restores known-good state from protected
storage. These capabilities, however, do not automatically produce selective
protection. The evaluated mechanisms trade operation coverage against
natural-workload specificity, or defer the decision to a trusted recovery
point with workload-dependent rollback cost.
Self-state defense should therefore be evaluated on three properties together:
coverage of the mutation surface, selectivity under representative legitimate
workloads, and recovery exposure when restoration is required. Our attack
space, contrastive workloads, and five-source measurement substrate provide a
reproducible basis for that evaluation. They also motivate specialized OS
defenses designed to turn host enforcement and evidence into selective
self-state decisions.

%% ===========================================================
%% References
%% ===========================================================
\bibliographystyle{plain}
\bibliography{references}

%% ===========================================================
%% Appendices
%% ===========================================================
\appendix

\section{Open Science}
\label{app:openscience}
The repository for all the code and data artifacts will be released with a permanent identifier (DOI) using Zenodo upon acceptance. Under \texttt{experiments/}, \texttt{code/} contains the implementations for
attack generation, telemetry collection, detection, prevention, and recovery,
while \texttt{results/} contains the frozen population and split manifests,
per-detector scored outputs, per-result SHA-256 inventories, and materialized
per-run self-state fixtures used by the detectors.

The original raw telemetry corpus---including syscall captures, provenance
graphs, large per-run event streams, and durable archival data---is excluded
because of its size. The released manifests retain the provenance metadata and
checksums associated with these inputs. The raw corpus is not required to
inspect or independently verify the reported calculations: the released result
JSON files and manifests contain the frozen derived outputs and population
definitions used to produce the paper's estimates. Recomputing those outputs
from the original telemetry requires access to the unreleased raw corpus.
Alternatively, reproducing the full pipeline from newly collected executions
requires privileged Linux collection infrastructure, isolated guest
environments, and evaluator-provided model credentials.

\section{Ethical Considerations}
\label{app:ethics}

Stakeholders include users and operators of self-hosted agents, framework and
model-service providers, open-source maintainers, and the security research
community. Potential harms include reuse of persistent-state attack techniques
against real deployments, unintended interaction with third-party systems, and
overgeneralization of results beyond the evaluated stack. The corresponding
benefit is reproducible evidence about where OS defenses provide effective
enforcement, attribution, and recovery, and where additional decision context
is required.

The study recruited no participants and collected no private user records. W1
and W2 derive from public task corpora, while W3 and W4 use authored scenarios.
The attack corpus uses inert markers and reserved \texttt{.invalid}
destinations; it does not deliver malware, access production accounts, or
target third-party systems. All trials operate on isolated experiment
workspaces using synthetic or public benchmark data.

For model-driven cases, operational landing and maliciousness adjudication are
kept separate: persistence of an induced marker does not by itself establish
maliciousness, which is determined through the pre-specified adjudication
procedure.

The released catalog contains ordinary file modifications, additions,
deletions, and permission changes. Potentially sensitive carrier text remains
restricted to the isolated corpus, and external-looking addresses use
non-routable placeholders. We judged the study and its publication ethically
justified because the experimental procedures avoid external state changes
while supporting reproducible evaluation of defensive mechanisms. Residual
misuse risk is further limited by inert payloads, explicit scope annotations,
and the absence of deployable malware, production credentials, or live
targets.

\section{Generative AI Usage}
\label{app:ai-usage}

Generative AI tools were used to assist with code implementation, prose
drafting, and consistency review. The authors selected the experimental
design, inspected the underlying artifacts, and adjudicated the scientific
claims. All AI-assisted text, code, and suggested references were reviewed and
validated by the authors as if authored directly by them; the authors remain
responsible for all submitted content and reported results.

\begin{table*}[t]
  \centering
  \caption{Execution environment for the five-source corpus. Resource values
  are the per-run limits enforced on the agent process tree, not the capacity
  of the physical host.}
  \label{tab:execution-environment}
  \footnotesize
  \begin{tabularx}{\textwidth}{@{}lX@{}}
    \toprule
    Component & Frozen setting \\
    \midrule
    Guest platform & Ubuntu 22.04.5 LTS; x86-64; glibc 2.35; Linux
      6.8.0-136-generic; ext4 root filesystem on a virtio block device;
      kernel BTF present \\
    Agent execution & Unprivileged runner UID 997; cgroup v2
      \texttt{cpu.max=80000 100000} (0.8 CPU quota) and
      \texttt{memory.max=805306368} (768 MiB) applied to the run process tree \\
    Clock discipline & UTC; NTP enabled and synchronized; wall-clock and
      monotonic anchors retained with each run \\
    File and syscall collection & inotify and fanotify Linux~6.8 UAPI;
      auditd/auditctl~3.0.7; a content-addressed custom libbpf eBPF collector;
      native SCAP capture with sysdig~0.41.4 \\
    Replay and enforcement tools & Falco~0.44.0; AIDE~0.19.3; Landlock ABI~4;
      nftables~1.0.2; Python~3.10.12 \\
    Run isolation & Fresh run workspace and state snapshot; one active capture
      per VM; attack runs confined to designated attack VMs \\
    \bottomrule
  \end{tabularx}
\end{table*}

\section{Artifact and Experimental Details}
\label{app:harness-setup}

\subsection{Execution Environment and Run Isolation}
\label{app:execution-environment}

The five-source experiments ran in x86-64 virtual machines with Ubuntu
22.04.5 LTS, Linux~6.8.0-136-generic, glibc~2.35, and an ext4 root filesystem.
Table~\ref{tab:execution-environment} records the environment parameters that
can affect observability, enforcement, or repeatability. Each run also retains
an environment fingerprint containing the kernel and tool versions, clock
state, runner identity, and hashes of the collector and eBPF binaries.

The supplemental prevention operation replay used a separate privileged
container on Linux~6.14.0-1015-nvidia, aarch64, and an ext4 bind mount. Controls
were installed by the supervisor, and each operation then ran as agent
UID~1000. AppArmor was enabled and enforcing, and the Landlock launcher was
compiled from the released source. This deterministic kernel probe is reported
as a separate experiment batch and is not pooled with the x86-64 five-source
corpus.

The resource envelope is enforced at the run boundary because it is the
resource allocation experienced by the agent and all descendants. Physical
host capacity is not an experimental variable. A second VM may contribute to
a result generation only after its kernel, collector, eBPF object, detector
stack, runner UID, and repository lineage match the frozen generation and a
clean-parity pilot passes. Captures are serialized within each VM so concurrent
experiments cannot contaminate event streams or drop accounting.

\subsection{Reference Adapter and Task Sources}
\label{app:harness-tasks}

The benchmark implements a generic state schema whose reference adapter binds
instruction roles to \texttt{SOUL.md}, \texttt{IDENTITY.md},
\texttt{USER.md}, \texttt{AGENTS.md}, and \texttt{TOOLS.md}; durable global
memory to \texttt{MEMORY.md}; episodic and topic-scoped memory to
\texttt{memory/*.md}; runtime configuration to \texttt{openclaw.json};
automation to \texttt{HEARTBEAT.md}; and capability bindings to
\texttt{credentials/.env}. The \texttt{openclaw-core} reference backend
implements the agent loop, tool dispatch, and file-backed state-loading
behavior exercised by the OpenClaw-oriented reference adapter using ordinary
Linux processes and files. Attack records retain the logical object ID and
role alongside the resolved target path.

W1 tasks are curated from the pinned Aider polyglot Python subset and use test
execution as their success criterion. W2 tasks are selected from FRAMES with
pinned source material and answer matching. W3 exercises self-configuration
and operations workflows using file-state assertions. W4 exercises general
assistant file workflows and is retained primarily for its natural host
activity. Corpus provenance, task identities, seeds, and success checks are
version controlled with the task definitions.

\subsection{Run Construction and Ground Truth}
\label{app:run-construction}

The unit of execution and detector decision is one run branch. For
model-driven attack cases, the harness constructs an attack branch and a
task-matched clean branch from the same task bundle, initial self-state, tool
environment, and model setting. Only the adversarial carrier differs. The
evaluated carriers are a poisoned user message, an ingested workspace file,
and tool-delivered content. Metadata and namespace attacks are instead
instantiated directly under the compromised agent principal, consistent with
the post-compromise threat model in Section~\ref{sec:ssa-def}. The separate
natural clean corpus samples the workload distribution without being
restricted to attack-matched tasks; it is split into natural-write and
clean-no-write outcomes according to whether the run changed a canonical
self-state object.

Every model-driven branch retains the initial state, the primary
post-execution snapshot, and a later persistence snapshot, named
\texttt{before\_a}, \texttt{after\_a}, and \texttt{after\_b}. The primary
mutation comparison is \texttt{before\_a} to \texttt{after\_a};
\texttt{after\_b} is a persistence diagnostic and is never substituted for the
primary outcome. File manifests record existence, type, mode, size, and
SHA-256 digest for canonical objects. For model-driven cases, operational
landing and semantic polarity are distinct labels: a marker shows that induced
content persisted, whereas the polarity review determines whether the adopted
state is malicious. Directly instantiated attacks are adjudicated from their
pre-specified malicious operation and observed post-state rather than from a
carrier-landing test.

Only valid executions adjudicated as malicious enter the attack-positive
denominator. For model-driven cases, successful persistent landing is
additionally required to establish end-to-end induction. Invalid,
execution-error, pending, and data-insufficient runs are retained with reasons
and are never converted to negative decisions.

Operation-observability validation cases form a separate measurement
population. Each validation case executes a deterministic operation under the
agent principal to test whether the collection substrate witnesses and
reconstructs the intended write, rename, unlink, or permission change. These
cases establish operation observability, not attack evidence, and do not enter
the 55-execution attack denominator.

\begin{table*}[t]
  \centering
  \caption{Complete structural-to-concrete catalog mapping. The 23 cells
  expand to 43 executable bindings.}
  \label{tab:catalog-inventory}
  \footnotesize
  \begin{tabularx}{\textwidth}{@{}l>{\raggedright\arraybackslash}Xr@{\hspace{2em}}l>{\raggedright\arraybackslash}Xr@{}}
    \toprule
    Cell & Concrete bindings & $N$ & Cell & Concrete bindings & $N$ \\
    \midrule
    Mem-M1-G1 & \texttt{MEMORY.md} & 1 & Inst-M1-G1 & \texttt{TOOLS.md}, \texttt{USER.md} & 2 \\
    Mem-M1-G2 & \texttt{MEMORY.md} & 1 & Inst-M1-G2 & \texttt{TOOLS.md}, \texttt{SOUL.md} & 2 \\
    Mem-M1-G3 & \texttt{MEMORY.md}, \texttt{memory/*.md} & 2 & Inst-M1-G3 & \texttt{USER.md}, \texttt{AGENTS.md} & 2 \\
    Mem-M1-G4 & \texttt{MEMORY.md}, \texttt{memory/*.md} & 2 & Inst-M1-G4 & \texttt{USER.md}, \texttt{IDENTITY.md} & 2 \\
    Mem-M2-G2 & \texttt{MEMORY.md}, \texttt{memory/*.md} & 2 & Inst-M2-G2 & \texttt{TOOLS.md}, \texttt{AGENTS.md} & 2 \\
    Mem-M2-G3 & \texttt{MEMORY.md}, \texttt{memory/*.md} & 2 & Inst-M2-G3 & \texttt{TOOLS.md}, \texttt{USER.md} & 2 \\
    Mem-M2-G4 & \texttt{MEMORY.md}, \texttt{memory/*.md} & 2 & Inst-M2-G4 & \texttt{SOUL.md} & 1 \\
    Mem-M3-G1 & \texttt{MEMORY.md}, \texttt{memory/*.md} & 2 & Inst-M3-G1 & \texttt{SOUL.md}, \texttt{AGENTS.md}, \texttt{IDENTITY.md} & 3 \\
    Mem-M4-G1 & \texttt{MEMORY.md}, \texttt{memory/*.md} & 2 & Inst-M4-G1 & \texttt{TOOLS.md}, \texttt{AGENTS.md} & 2 \\
    \addlinespace[2pt]
    Cfg-M1-G1 & \texttt{openclaw.json}, \texttt{HEARTBEAT.md} & 2 & Cfg-M1-G4 & \texttt{openclaw.json}, \texttt{HEARTBEAT.md} & 2 \\
    Cfg-M1-G2 & \texttt{openclaw.json} & 1 & Cfg-M4-G1 & \texttt{openclaw.json}, \texttt{HEARTBEAT.md} & 2 \\
    Cfg-M1-G3 & \texttt{openclaw.json}, \texttt{HEARTBEAT.md} & 2 & & & \\
    \bottomrule
  \end{tabularx}
\end{table*}

\subsection{Telemetry Collection and Admission Criteria}
\label{app:collection-admission}

Each run retains raw inotify, fanotify, audit, eBPF, and SCAP streams together
with source-health records, snapshots, the task and state manifests, and the
environment fingerprint. The normalizer joins subject identity, process
lineage, operation, resolved object, outcome, and evidence provenance. Inotify
remains an independent chronology and overflow check; the joined graph uses
fanotify, audit, eBPF, and libsinsp-reconstructed SCAP evidence. Detector
exporters then select the substrate native to each baseline. Thus, five-source
collection is a measurement-integrity property, not a claim that every
detector consumes all five streams.

A run is admitted only if required streams are nonempty, source drop and
overflow counters are zero, SCAP termination is valid, and the effective
file-descriptor-to-path resolution rate is at least 0.95 with zero excluded
writes. The bridge also requires the controlled runner UID and an evaluable
provenance graph. Observation-generation and derivation-generation identifiers,
input hashes, commands, tool exits, and output hashes are retained so that
cross-generation mixing can be rejected rather than silently scored.

\begin{table}[t]
  \centering
  \caption{Detector evaluation protocols. A decision is made per execution;
  native event counts and scores are retained as secondary outputs.}
  \label{tab:baseline-protocols}
  \footnotesize
  \begin{tabularx}{\columnwidth}{@{}>{\raggedright\arraybackslash}p{.32\columnwidth}X@{}}
    \toprule
    Method and input & Positive decision \\
    \midrule
    AIDE~0.19.3; snapshots & At least one added, removed, or changed canonical
      object in \texttt{after\_a} \\
    Falco~0.44.0; SCAP & At least one benchmark-specific canonical-state
      mutation event by UID 997 or a descendant \\
    STIDE; syscall sequences & At least one unseen 6-gram in a process from the
      pre-specified profile-specific core executable set \\
    UNICORN~\cite{han2020unicorn}; role-typed provenance graphs & Official
      analyzer returns abnormal at the pre-specified 3.0-standard-deviation
      threshold in the profile-conditioned model \\
    B1/B2; size and timing & At least one resolved self-state write with
      \(\mathrm{score}(e;k)\geq2\); B1 uses pooled and B2 profile-specific
      clean estimates \\
    \(L_1\), CART, FIGS; size/timing features & Model score reaches the
      training-only maximum-Youden-\(J\) threshold \\
    \bottomrule
  \end{tabularx}
\end{table}

\subsection{Structural Catalog}
\label{app:catalog-mapping}

The 23-cell matrix is the valid Target $\times$ Mechanism $\times$
Granularity projection after fixing the Temporal axis to T1. It expands to 43
catalog entries because a structural cell can have multiple concrete target
files. T2 through T4 remain taxonomy dimensions and are not counted as
empirical matrix coverage in this paper.

MITRE ATLAS supplies AI-state anchors, including context poisoning and agent
configuration modification; MITRE ATT\&CK supplies stored-data manipulation,
file deletion, and permission-modification anchors; OWASP Agentic AI supplies
the agent-level consequence categories. These mappings establish that the
matrix instantiates established threat concepts, but they are not themselves
empirical attack observations.

Table~\ref{tab:catalog-inventory} gives the complete mapping promised by the
23-cell abstraction. Catalog IDs concatenate the abbreviated target,
mechanism, and granularity, for example, \texttt{Mem-M1-G1}. M1 modifies
existing state, M2 adds state, M3 deletes an object, and M4 denies access;
G1 through G4 denote whole-object, large, line-scale, and minimal changes.
Multiple bindings within a cell increase within-cell heterogeneity without
increasing structural coverage.

The released executable catalog records each operation's attack ID, target
file, operation type, before/after size and hash, and success status. The
realized attack population uses two execution modes: model-driven sessions for
persistent M1/M2 mutations and direct post-compromise instantiation for the
remaining metadata or namespace mutations. The execution mode records whether
an attack additionally demonstrates end-to-end model inducibility; it does not
change its status as an attack execution in the OS-level evaluation.
Operation-observability validation cases are recorded separately and are not
counted as attack executions.

\begin{table*}[t]
  \centering
  \caption{Confidence intervals and natural-control decomposition for
  Table~\ref{tab:fpr-legitimate}. Intervals are 95\% cluster-bootstrap
  intervals.}
  \label{tab:detector-details}
  \footnotesize
  \setlength{\tabcolsep}{6pt}
  \begin{tabularx}{\textwidth}{@{}l c c
    >{\centering\arraybackslash}X@{}}
    \toprule
    \textbf{Method} & \textbf{Attack TPR interval (\%)} &
    \textbf{Natural FPR interval (\%)} & \textbf{Write / no-write} \\
    \midrule
    AIDE & 87.5--100 & 33.3--73.3 & 32/32 / 0/28 \\
    Falco & 66.7--88.7 & 20--48.3 & 21/32 / 0/28 \\
    STIDE & 100 & 91.7--100 & 32/32 / 26/28 \\
    UNICORN & 73.2--95.1 & 53.7--91.5 & 27/27 / 8/20 \\
    \midrule
    B1 pooled & 59.3--100 & 5--31.7 & 10/32 / 0/28 \\
    B2 per-profile & 70.8--100 & 0--26.7 & 7/32 / 0/28 \\
    \(L_1\) logistic regression & 41.7--90 & 5--20 & 7/32 / 0/28 \\
    CART & 41.9--90 & 0--13.3 & 3/32 / 0/28 \\
    FIGS & 34.8--80 & 0--13.3 & 3/32 / 0/28 \\
    \bottomrule
  \end{tabularx}
\end{table*}

\subsection{Pre-Specified Baseline Decision Rules}
\label{app:baseline-rules}

Table~\ref{tab:baseline-protocols} separates each implementation from the
benchmark-specific input adapter and positive decision. The AIDE policy and
Falco rule are intentionally scoped to canonical self-state; they are not
advertised as vendor-default configurations. No held-out natural execution
selects a feature, executable set, rule, or decision threshold.

AIDE uses the policy
\texttt{p+i+n+u+g+s+m+c+sha256}. A method is evaluable only when its required
inputs and execution checks succeed. AIDE requires successful initialization
and a no-change control check; Falco requires valid replay and canonical-path
qualification. STIDE requires a trained core model and a process instance of
at least 106 sequence-eligible syscalls. UNICORN requires successful graph
parsing and analyzer completion, yielding 41/55 evaluable attack graphs and
47/60 evaluable natural graphs. B1, B2, and the supervised methods require a
resolved self-state write with valid size features, which defines 23/55 attack
executions. The remaining 32 attack executions are reported as outside the
method definition and are never converted into negative predictions. On the
natural corpus, an execution with no self-state write produces no candidate
write event and therefore no positive execution-level decision; this does not
expand the attack-side operation coverage of the write-specific methods.

The supervised feature vector contains signed and absolute size change,
before/after size, write count, inter-write timing and span, timing
availability, state layer, workload profile, and object bucket. The 23 attacks
are partitioned into five fixed, approximately profile-stratified folds with
all executions from one scenario kept together. Each outer model trains on the
remaining attack folds and all 176 clean training executions; a grouped inner
five-fold procedure selects the maximum-Youden-\(J\) threshold. For natural
FPR, the model is refit on all 23 attacks and the clean training corpus, then
applied once to the independent natural test corpus. Attack-fold scores and
separately fitted natural-test scores are not pooled into a single
area-under-curve statistic.

For B1 and B2, an unseen key receives an infinite anomaly score (represented
as \(10^6\) in the implementation). If a timing baseline has zero variance,
an exact match contributes zero and a mismatch contributes five. If timing is
unavailable, the detector uses only the size term in
Equation~\ref{eq:anomaly-score}.

For every method, natural FPR is the fraction of positive held-out natural
executions. Attack TPR is the fraction of positive attack executions on the
method's defined population and is reported alongside attack coverage. Point
estimates use executions. Confidence intervals resample independent clean
scenarios or attack folds while retaining correlated executions, and boundary
cells additionally report cluster-any Wilson intervals. Natural-write and
clean-no-write controls are reported separately.
Table~\ref{tab:detector-details} gives the complete inference details behind
the primary comparison. UNICORN remains conditional because graph-processing
failures are data-insufficient rather than negative decisions.

\begin{table}[t]
  \centering
  \caption{Distinct self-state object buckets modified per clean session. The
  profile rows are descriptive contrasts, not independent deployment samples.}
  \label{tab:rollback-exposure}
  \footnotesize
  \setlength{\tabcolsep}{4pt}
  \begin{tabular}{@{}lrrrrr@{}}
    \toprule
    \textbf{Profile} & \textbf{$n$} & \textbf{Mean} & \textbf{Median} &
    \textbf{90th pct.} & \textbf{Max} \\
    \midrule
    W1 & 60 & 0.15 & 0 & 1 & 1 \\
    W2 & 60 & 0.83 & 1 & 1 & 1 \\
    W3 & 66 & 2.09 & 2 & 3 & 3 \\
    W4 & 50 & 0.74 & 0 & 2 & 3 \\
    \midrule
    All & 236 & 0.99 & 1 & 2 & 3 \\
    \bottomrule
  \end{tabular}
\end{table}

\subsection{Evaluation Lineage and Denominators}

The clean training population contains 176 executions from 81 scenarios. The
held-out natural population contains 60 executions from 20 disjoint scenarios,
with three replicates per scenario. Its membership was fixed before scoring and
was not expanded after observing detector outputs. Together, these 236 clean
executions are also used for the descriptive recovery-exposure analysis in
Table~\ref{tab:rollback-exposure}; the held-out subset remains reserved for
detector false-positive evaluation. The five-source collector retains inotify,
fanotify, audit, eBPF, and SCAP data, while detector-specific exporters select
only the substrate each baseline consumes.

The attack population contains 55 adjudicated malicious executions grouped
into 52 attack folds. Thirty-five involve model-driven persistent writes, and
20 instantiate metadata or namespace mutations directly under the compromised
agent principal. Both groups enter the OS-level attack denominator; their
distinction limits only claims about end-to-end model inducibility. The
size-and-timing subset contains 23 attack executions in 20 scenarios. The
supervised outer split partitions these scenarios into five fixed random folds.
The 21 task-matched clean branches used for provenance analysis are excluded
from detector training and retained only for the secondary paired comparison.
Superseded development populations and leaky clean comparisons remain in the
artifact for provenance but are not paper estimates.

\subsection{Prevention Replay and Recovery-Cost Supplement}
\label{app:prevention-replay}

The prevention replay executes each control--operation cell from a freshly
initialized \texttt{MEMORY.md}. Success is determined from the requested
post-state effect, not merely the absence of an error return: overwrite and
rename-over must produce the marker bytes, truncate must produce the requested
size, unlink must remove the path, and \texttt{chmod/write} must both reset the
mode and land the overwrite. The raw result records the pre-state, syscall
outcome, kernel error code, and post-state for every cell. The paired
direct-write probe is retained separately because it directly compares an
attack-designated write with a corresponding operator-approved write under the
same agent principal and policy.

Table~\ref{tab:rollback-exposure} characterizes the clean write process used to
interpret recovery cost. Counts use the same successful canonical-write
predicate as the detector substrate and aggregate paths into canonical
self-state object buckets.

Of the 236 executions, 150 modify at least one self-state object and 86 modify
none. Among writing sessions, the mean is 1.56 objects, the median is 1, the
90th percentile is 2, and the maximum is 3. The corresponding write-event
counts over all sessions have mean 1.28, median 1, 90th percentile 3, and
maximum 4. We do not multiply these session summaries by a longer backup
interval: doing so would count object-update incidences and can double-count a
path modified in multiple sessions, whereas the recovery metric counts
distinct changed paths.

\subsection{Known Scope Limitations}

There are several scope boundaries to be highlighted for the sake of clarity. The artifact has one live backend. Alternate layout
testing demonstrates rebinding, not cross-system generality. The four profiles
are contrastive rather than a deployment sample, and five held-out scenarios
per profile make profile-level FPR descriptive. Model-driven attack evidence is
concentrated in M1/M2 persistent writes, while the remaining metadata and
namespace attacks are instantiated directly under the compromised agent
principal; this distinction limits claims about model inducibility rather than
their inclusion in the OS-level attack evaluation. Prevention replays six
operations against five configurations on one durable-memory object  and does not
a proof over all Linux authority paths or policy designs. Recovery covers the
repository-isolation matrix, eight protected attacks, and a descriptive
236-session cost analysis, not arbitrary recovery points or a measured
multi-session distinct-path curve.

\end{document}